\documentclass[twocolumn,preprintnumbers,amsmath,prd,amssymb,superscriptaddress]{revtex4}

\usepackage{capt-of,graphicx}% Include figure files
\usepackage{dcolumn}% Align table columns on decimal point
\usepackage{bm}% bold math
\usepackage{hyperref}
\usepackage{color}

\def\sa12{\rm{SA_{12}}}

\begin{document}

%\preprint{AS-TEXONO/12-??}

\title{Design and Performance of a Hybrid Fast and Thermal Neutron Detector}

%Begin Author List
\author{ M.~K.~Singh }  \altaffiliation[Corresponding Author: ]{ \corrmks }\affiliation{ \1 } \affiliation{ \2 } 
\author{ A.~Sonay }  \affiliation{ \1 } \affiliation{ \3 }
\author{ M.~Deniz }  \altaffiliation[Corresponding Author: ]{ \corrmz } \affiliation{ \1 } \affiliation{ \3 } \affiliation{ \4 }
\author{ M.~Agartioglu }  \affiliation{ \1 } \affiliation{ \3 }
\author{ G.~Asryan }  \affiliation{ \1 } 
\author{ G.~Kiran~Kumar }  \altaffiliation[Present Address: ]{ \kiran } \affiliation{ \1 } 
\author{H.~B.~Li }  \affiliation{ \1 } 
\author{ J.~Li }  \affiliation{ \5 }
\author{ F.~K.~Lin }  \affiliation{ \1 } 
\author{ S.~T.~Lin }  \affiliation{ \1 } \affiliation{ \6 }
\author{ V.~Sharma }  \affiliation{ \1 } \affiliation{ \2 }
\author{ L.~Singh }  \affiliation{ \1 } \affiliation{ \2 }
\author{ V.~Singh }  \affiliation{ \2 } 
\author{ V.~S.~Subrahmanyam }  \affiliation{ \2 } 
\author{ A.~K.~Soma }  \altaffiliation[Present Address: ]{ \arun } \affiliation{ \1 } \affiliation{ \2 }
\author{ H.T.~Wong }  \altaffiliation[Corresponding Author: ]{ \corrhtwong } \affiliation{ \1 }
\author{ S.~W.~Yang }  \affiliation{ \1 } 
\author{ I.~O.~Yildirim }  \affiliation{ \1 } \affiliation{ \4 }
\author{ Q.~Yue }  \affiliation{ \5 }

\newcommand{\1}{Institute of Physics, Academia Sinica, Taipei 11529, Taiwan.}
\newcommand{\2}{Department of Physics, Institute of Science, Banaras Hindu University, Varanasi 221005, India.}
\newcommand{\3}{Department of Physics, Dokuz Eyl\"{u}l University, Buca, \.{I}zmir TR35160, Turkey.}
\newcommand{\4}{Department of Physics, Middle East Technical University, Ankara TR06531, Turkey.}
\newcommand{\5}{Department of Engineering Physics, Tsinghua University, Beijing 100084, China.}
\newcommand{\6}{Department of Physics, Sichuan University, Chengdu 610065, China.}

\newcommand{\corrmks}{singhmanoj59@gmail.com}
\newcommand{\corrmz}{muhammed.deniz@deu.edu.tr}
\newcommand{\corrhtwong}{htwong@phys.sinica.edu.tw}

\newcommand{\arun}{Department of Physics, University of South Dakota, Vermillion, SD 57069, USA.}
\newcommand{\kiran}{Physics Department, KL University, Guntur 522502, India.}

\collaboration{The TEXONO Collaboration}

%%\noaffiliation
%%End Author List


% PACS, the Physics and Astronomy % Classification Scheme.
%\pacs{
%%95.35.+d,
%%29.40.-n,
%%98.70.Vc
%}
%Use showkeys class option if keyword %display desired
%%\keywords{
%%Dark matter,
%%Radiation Detector,
%%Background radiation
%%}

\begin{abstract}
We report the performance and characterization of a custom-built hybrid detector consisting of BC501A liquid scintillator for fast neutrons and BC702 scintillator for thermal neutrons. The calibration and the resolution of the BC501A liquid scintillator detector are performed. The event identification via Pulse Shape Discrimination (PSD) technique is developed in order to distinguish gamma, fast and thermal neutrons. Monte Carlo simulation packages are developed in GEANT4 to obtain actual neutron energy spectrum from the measured recoil spectrum. The developed methods are tested by reconstruction of $^{241}$AmBe($\alpha$, n) neutron spectrum.

\end{abstract}

\maketitle

\section{Introduction and Physics Motivations}
Experiments in neutrino physics \cite{neutrino} and dark matter searches \cite{dm} often involve very small interaction rates, and therefore desire a good detection sensitivity of low count rates of background. The elimination of known background signals from the physical signals constitutes a major part of the studies. However, complete reduction of background, especially neutron origin signals, is not possible even for very deep underground laboratories. Therefore, direct measurement of neutron background and its elimination are very crucial for such experiments. 

With these as objectives and within the framework of the TEXONO research program \cite{texono}, we develop some methods to construct actual neutron energy spectra and open a new research window to determine the neutron background from the direct measurement of neutron flux at the experimental site where the physical data taking will require low energy and low background environments, in order to determine its contribution to the background and eliminate it from the physical signals.

The "hybrid neutron detector" (HND) custom-built for this study consists of Bicron BC501A \cite{bc501a}, also known as NE231 or EJ301, liquid scintillator detector, which is sensitive for fast neutrons and BC702 \cite{bc702} organic scintillator detector, which is sensitive to thermal neutrons~\cite{bicron}. Liquid and organic scintillator detectors are widely used for neutron detection due to their high detection and capture efficiency. However, they are also sensitive to gamma rays as well. The $n/\gamma$ discrimination can be performed effectively with these scintillator detectors via Pulse Shape Discrimination (PSD) technique due to the different behavior of decay time component of the pulse shape. One of the purposes of this study is to develop a PSD technique providing a good and effective discrimination of neutron against gamma background and thereby construct actual neutron energy spectrum from its measured recoil spectra.

Fast neutrons can be produced by cosmic-ray interaction with environmental or surrounding materials as a secondary products or by spontaneous fission products via $(\alpha,n)$ reactions coming from radioactive materials in the rock such as naturally occurring isotopes of $^{238}U$ and $^{232}Th$ decay chains. The main physical mechanism of fast neutron detection is that the incident fast neutron scatters from the protons in the scintillating medium until it leaves from the detector with partial energy deposition or is totally captured. The scintillation signal is produced during the interaction by each recoiling proton. On the other hand, in case of incident thermal neutron, it diffuses in the detector medium and captured by $^{6}Li$ in a fine ZnS(Ag) phosphor powder via nuclear reaction of $^{6}Li(n,\alpha)^{3}H$ producing scintillations by the resulting alpha particle and triton. The reason of using $^{6}Li$ as the dopant is that it has the advantage of high neutron capture probability and produces two energetic charged particles that create scintillation.

The structure of the paper is as follows: Design and Construction of the HND will be discussed in Section-II. Data taking and detector performance (the DAQ system, energy calibration, event identification and PSD techniques) will be discussed in Section-III. Reconstruction of actual neutron spectra via Doroshenko and Gravel unfolding methods will be discussed in Section-IV.

\begin{figure}[hbt]
\includegraphics[width=7cm]{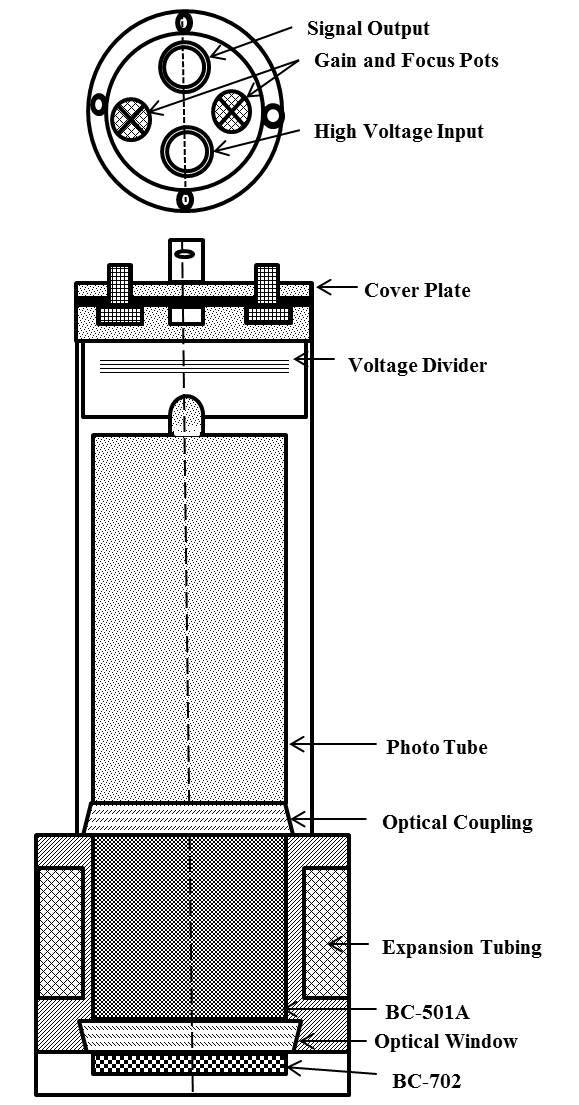}

\caption{Schematic diagram of the HND.}
\label{fig::detector}
\end{figure}

\section{Design and Construction}

\subsection{Detector Structure}
Neutron detector used in this study has a hybrid structure, bringing two different types of target materials to operate at the same time, Bicron BC-501A liquid scintillator having 0.113 liter cell volume and BC-702 type scintillator enriched to 95\% $^6$Li in a fine ZnS(Ag) phosphor powder. The scintillation light output is readout by Hamamatsu photomultiplier tube (PMT). The schematic diagram of the detector is shown in Fig.~\ref{fig::detector}. 

The selected detector dimension is similar to those of germanium detector cryostat used in experiments at the Kuo-Sheng Reactor Neutrino Laboratory (KSNL) \cite{ksnl} and China Jinping Underground Laboratory (CJPL) \cite{cjpl}. Therefore, HND can be installed, replacing the Ge-target, within the well of an NaI(Tl) Anti-Compton detectors under the same shielding configurations as those experiments to provide measurements of the ambient neutron background.

BC-501A is sensitive for the fast neutron detection while BC702 has high efficiency of detection of thermal neutrons. Both scintillator detectors have good fast time response as well as the pulse shape discrimination property, which enables isolation of the gamma events (due to different signal characteristics for proton and electron recoil events, enabling to distinguish neutron hit events from those of gammas). Therefore this HND provides good discrimination against gamma background \cite{psd1}. There are a large number of PSD studies for neutron detectors available in the literature \cite{psd2}.
\begin{figure}[hbt]
\includegraphics[width=8cm]{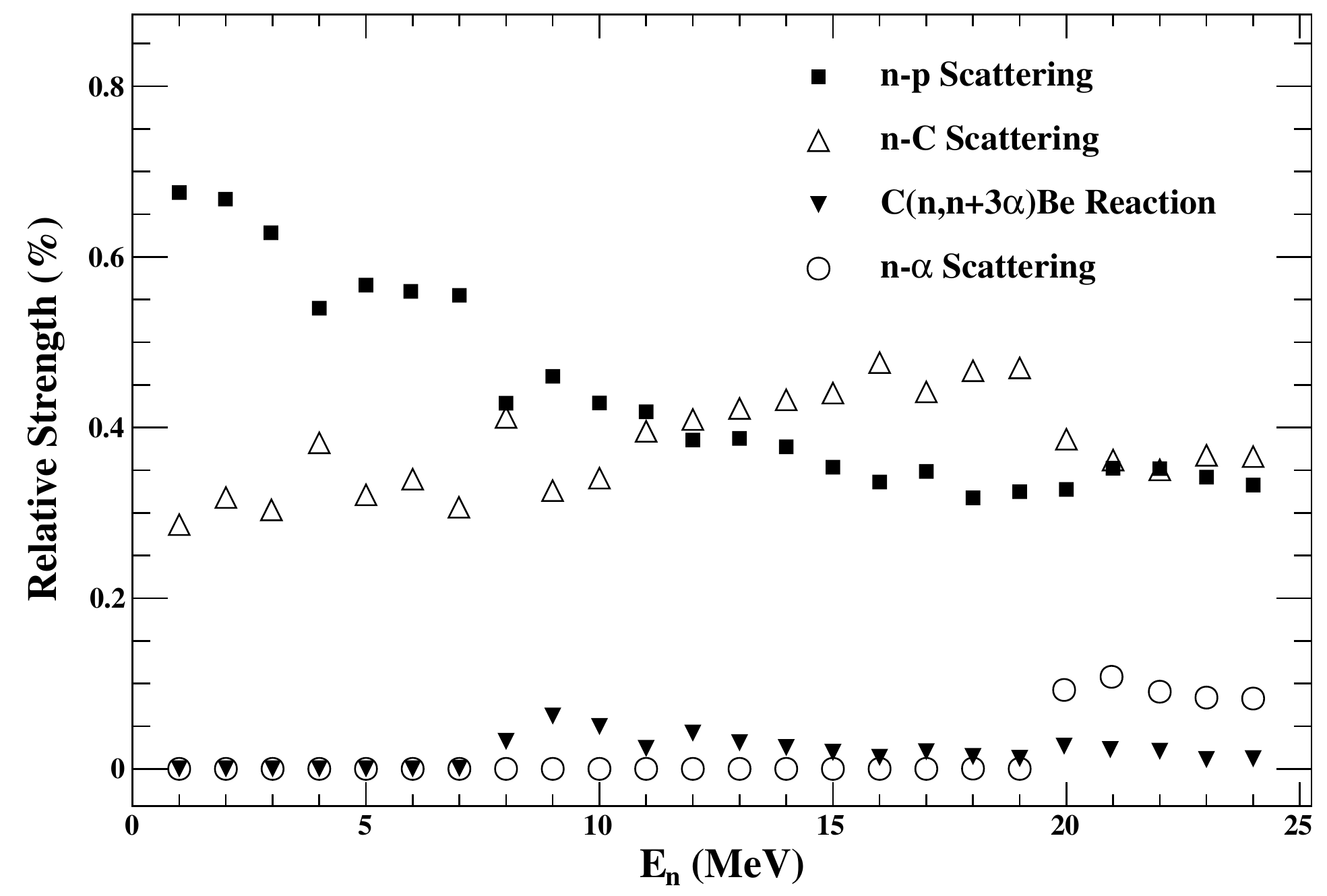}
\caption{Geant4 simulations for the proportions of the reactions occurred in the BC501A neutron detector.}
\label{fig::g4int}
\end{figure}

\begin{figure}
%\begin{center}
\textbf{(a)}\\
\includegraphics[width=7.8cm]{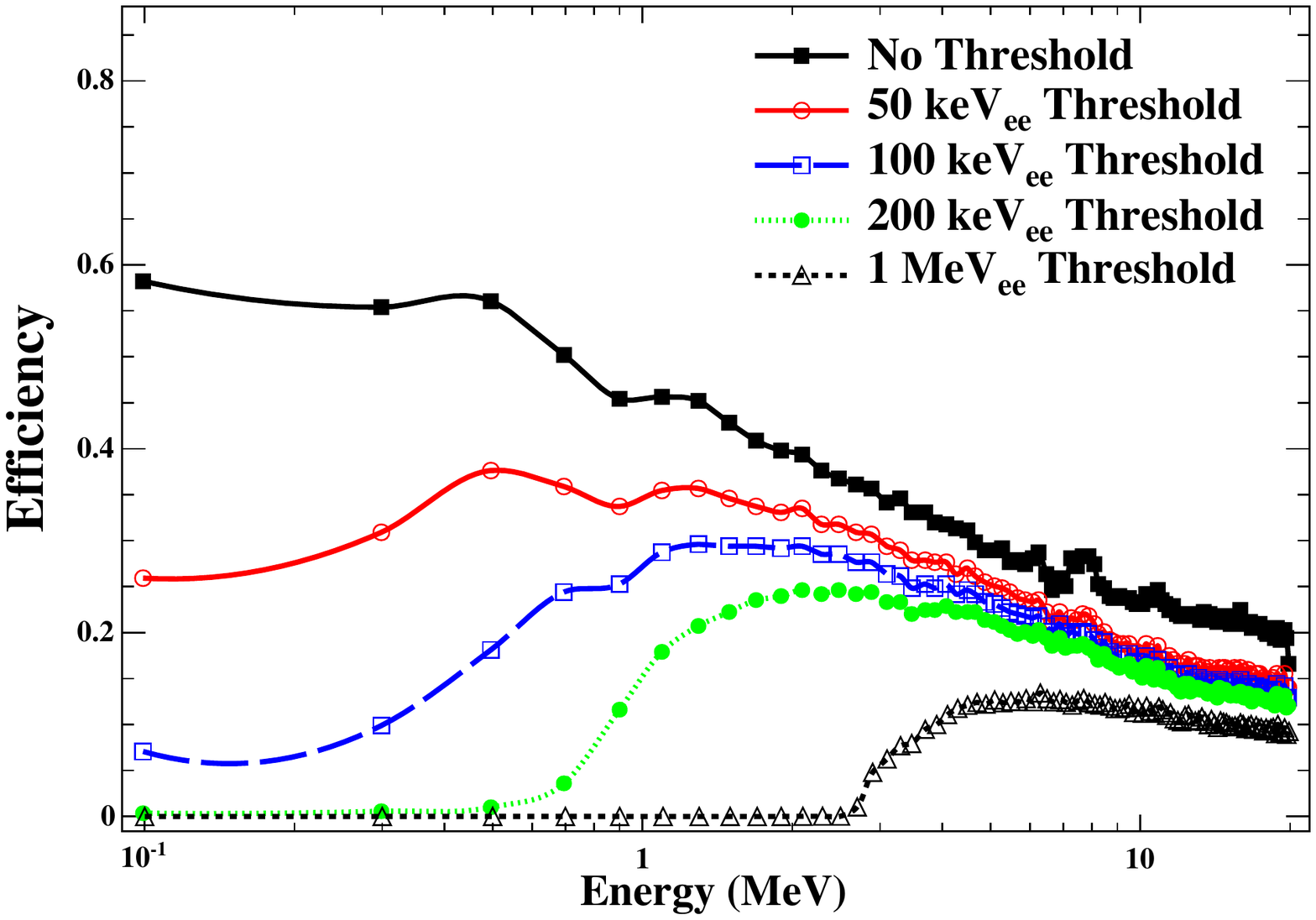}\\
\textbf{(b)}\\
\includegraphics[width=7.8cm]{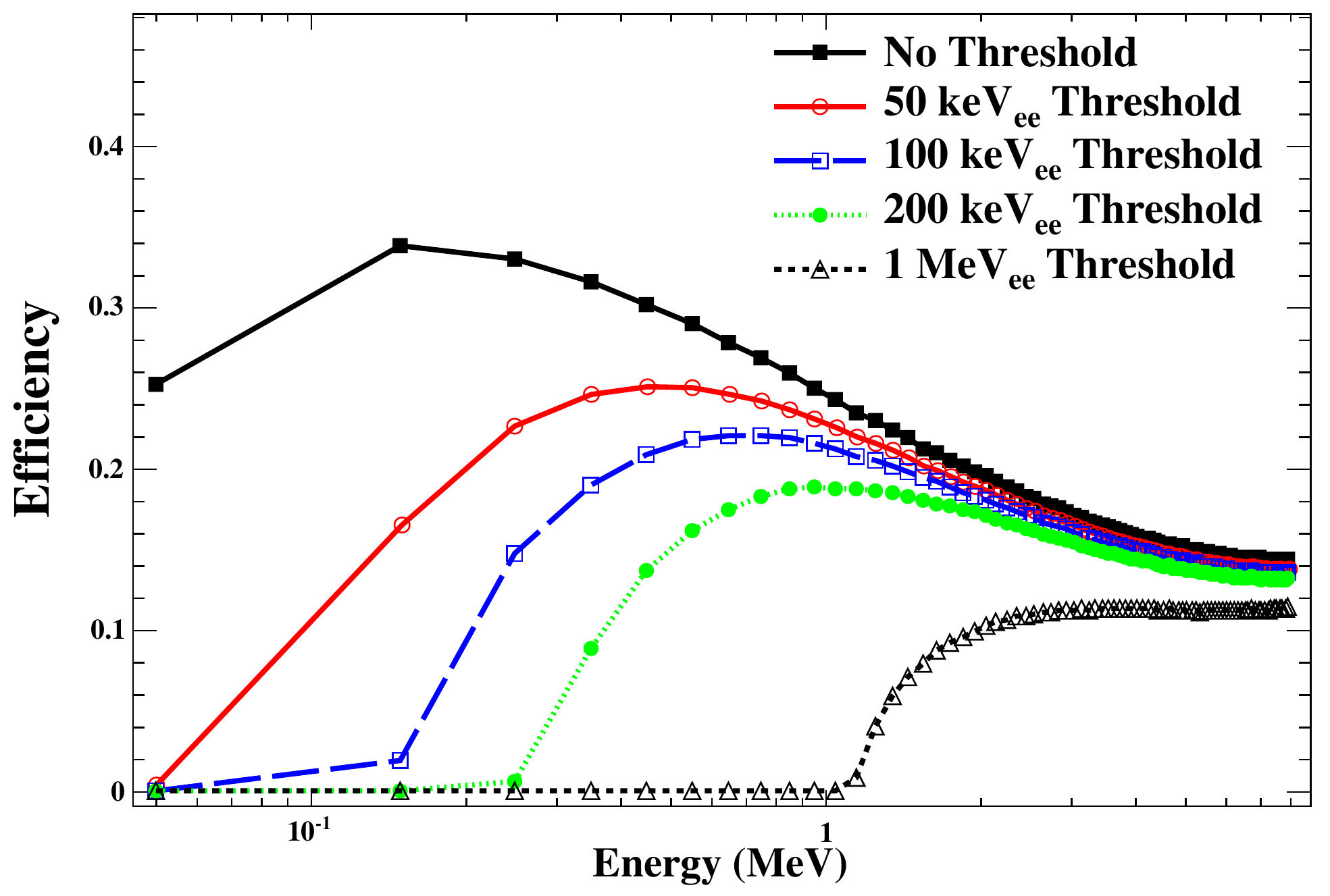}\\
\textbf{(c)}\\
\includegraphics[width=7.8cm]{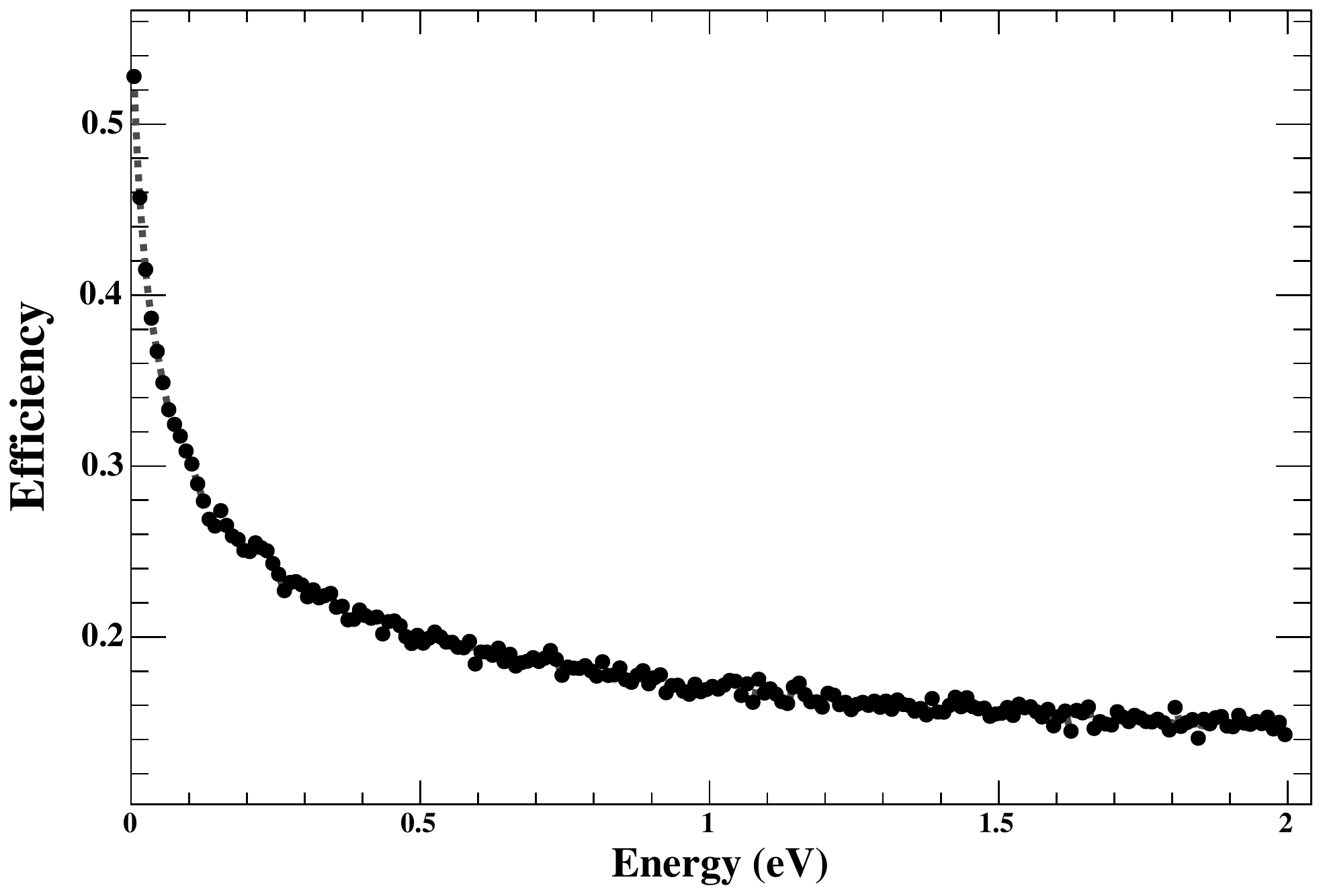}
\caption{(color online) Simulated efficiency of (a) BC501A with respect to incident neutron energy, (b) BC501A with respect to incident gamma energy. Different curves belong to different threshold. Uppermost curve is for no threshold and from top to bottom, at threshold values of 50 $keV_{ee}$, 100 $keV_{ee}$, 200 $keV_{ee}$ and 1 $MeV_{ee}$, (c) BC702 at several slow neutron energies.}
\label{fig::effND}
\end{figure}
\subsection{Physical Interactions}

BC-501A liquid scintillator detector is designed to yield good PSD discrimination against gamma and neutron incident particles. BC-501A is an organic liquid scintillator containing 4.82$\times10^{22}$ and 3.98$\times10^{22}$ atoms of hydrogen and carbon per cm$^{3}$, respectively. High density of hydrogen and carbon atoms present in its compound makes it a good target material for detection of fast neutrons in the $MeV$ range via the reactions of,
\begin{eqnarray}
%\begin{multline}
 p(n,n)p, \nonumber \\
^{12}C(n,n)^{12}C, \nonumber \\
^{12}C(n,n+3 \alpha),  \nonumber \\
^{12}C(n,\alpha)^{9}Be.
\label{eq::ndint}
%\end{multline}
\end{eqnarray}
A Geant4 Monte Carlo Simulation for the proportions of the reactions listed in Eq.~\ref{eq::ndint} are illustrated in Fig.~\ref{fig::g4int}. As shown, the most dominant interactions are $p(n,n)p$ and $^{12}C(n,n)^{12}C$. As the incident neutron energy increases the proportion of the $^{12}C(n,n)^{12}C$ interaction increases and becomes dominant above around 10 $MeV$. However, the regular neutron energy range in underground labs is usually below 10 $MeV$~\cite{tomasello}. Therefore, $p(n,n)p$ interaction is still the most probable and dominant interaction among the neutron detection channels for underground experiments.

BC702 detector is designed as a 6.35 mm thick and 50.8 mm diameter disc. BC702 is sensitive for slow/thermal neutron detection, which are present in the background environment mostly as a result of moderation of the fast neutrons via elastic scattering in the shielding and other materials. BC702 is highly efficient such that detection efficiency of neutrons with kinetic energy around 0.01 eV is above 50\% and this value rapidly decreases above 0.1 eV. BC702 scintillator is composed of 11 mg of $^6$Li per cm$^3$ with 95\% purity which is dispersed in ZnS(Ag) phosphor powder. The detector provides a good capture efficiency for thermal neutrons due to large neutron capture cross-section of $^{6}Li$, the reaction can be written as,
\begin{equation}
^{6}Li + n \rightarrow t~(2.05~MeV) + \alpha~(2.73~MeV)\label{eq::li}.
\end{equation}

The detection mechanism is neutron absorption by $^6$Li is given in Eq.~\ref{eq::li} where the resulting $\alpha$ particle and triton with recoil kinetic energy induces scintillation light upon their interaction with ZnS(Ag). BC702 detector provides very good discrimination of thermal neutrons against both gamma and fast neutron background.

The simulated efficiencies of the detector for fast neutrons, gamma and slow neutrons are illustrated in Figure~\ref{fig::effND}.
\begin{table}[hbt!]
\centering
\caption{Quenching factor parameters for BC501A liquid scintillator for proton and alpha particles.} \label{tab::cecil}
\begin{ruledtabular}
\begin{tabular}{l*5 cc}
Particle & A1 & A2 & A3 & A4 \\ \hline
p        & 0.83 & 2.82 & 0.25 & 0.93 \\
$\alpha$ & 0.41 & 5.9 & 0.065 & 1.01 \\
\end{tabular}
\end{ruledtabular}
\end{table}
\begin{figure}[hbt]
\includegraphics[width=8cm]{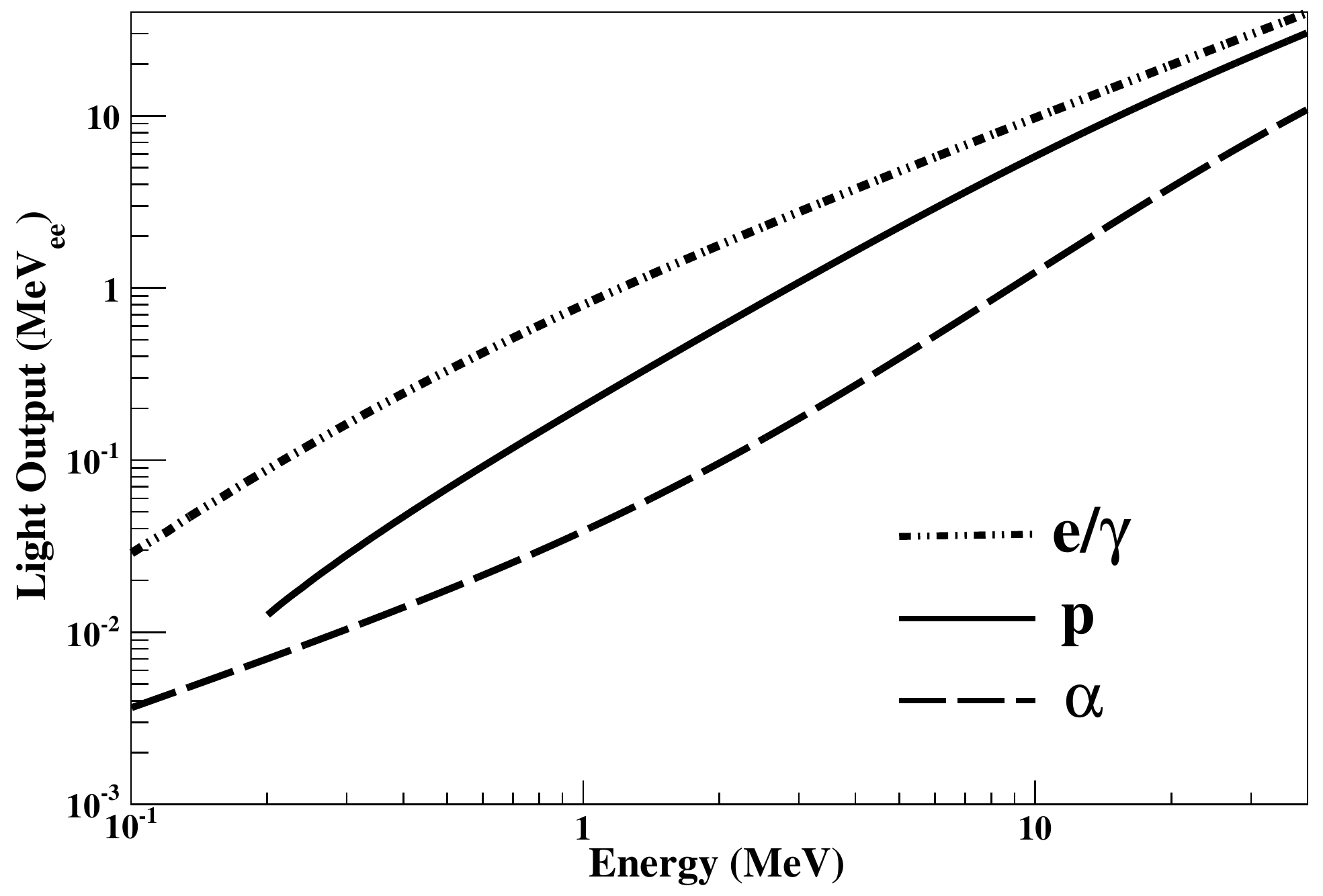}
\caption{Simulated light outputs for $e/\gamma$, proton and alpha particle versus incident energy.}
\label{fig::quench}
\end{figure}

\subsection{Quenching Effect}

The scintillating photons can be produced via photoelectric effect, Compton scattering or pair production, while neutrons are scattered inside the scintillator transferring their energies via the mechanisms given in Eq.~\ref{eq::ndint}. The energies due to neutron interactions are quenched compared to those from $e/\gamma$ interactions since the particles causing scintillation are charged and heavy. Therefore, simulated neutron recoil energy spectra has to be quenched to do comparison with the experimental measurements.

The quenched light output can be obtained by Eq.3 as,
\begin{equation}
L(E) = A_{1}\times E-A_{2} \times (1-e^{-A_{3} \times E^{A_{4}}}),
\label{eq::que}
\end{equation}
where $A_{1},~A_{2},~A_{3}$ and $A_{4}$ are quenching factor parameters listed in Table~\ref{tab::cecil} for BC501A liquid scintillator
detector~\cite{cecil}, $E$ is the recoil energy before quenched, $L(E)$ is the quenched energy which can be called as light output. The light output of electron, proton and alpha particle are illustrated in Figure~\ref{fig::quench}.
\begin{figure}[hbt]
\includegraphics[width=7cm]{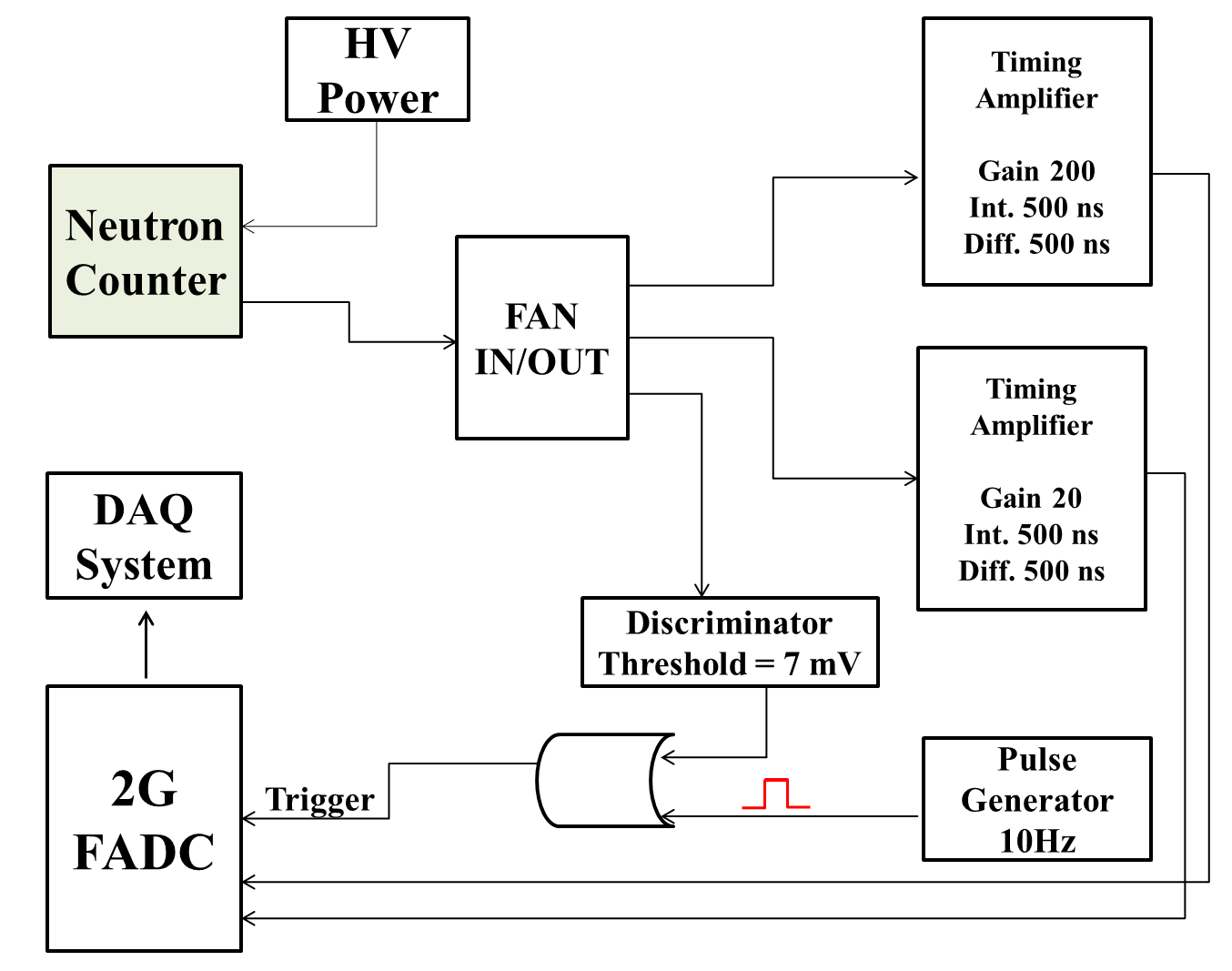}

\caption{The schematic block diagram of data acquisition (DAQ) electronic readout setup.}
\label{fig::readout}
\end{figure}

\section{Data Taking and Detector Performance}

The TEXONO experiment at KSNL has adopted PXI DAQ system running in Windows environment with Labview package program. The data is taken by this new DAQ system with Fast Timing Amplifiers (TAs). The signal from the anode of the PMT is directly connected to FAN IN/OUT to make two identical outputs, these two outputs are connected to two TAs with gains of 20 and 200 for the high energy and the low energy settings, respectively, while the integration and differentiation time for both TAs are same, which is 500 ns. The fast signals after TAs are recorded by 8 bit dynamic range of Flash Analog to Digital Converter (FADC) running at 2 GHz sampling rate. With this configuration we are able to cover low energy as well as high energy spectra at the same time. The schematic block diagram of data acquisition (DAQ) electronic readout setup is shown in Fig.~\ref{fig::readout}.
\begin{table}[hbt!]
\centering
\caption{\label{tab::compt} The list of gamma sources and their Compton Edge energies that are used in the calibration of BC-501A liquid scintillator neutron detector.}

\begin{ruledtabular}
\begin{tabular}{l*4 cc}
Source              & $E_{\gamma}(MeV)$ & $E_{c}(MeV_{ee})$ \\
\hline
${}^{22}Na$         & 0.511, 1.274     & 0.341, 1.062    \\
${}^{137}Cs$        & 0.662            & 0.478           \\
${}^{60}Co$         & 1.173, 1.332     & 0.963, 1.120    \\
\end{tabular}
\end{ruledtabular}
\end{table}
\begin{figure}[hbt]
\includegraphics[width=8cm]{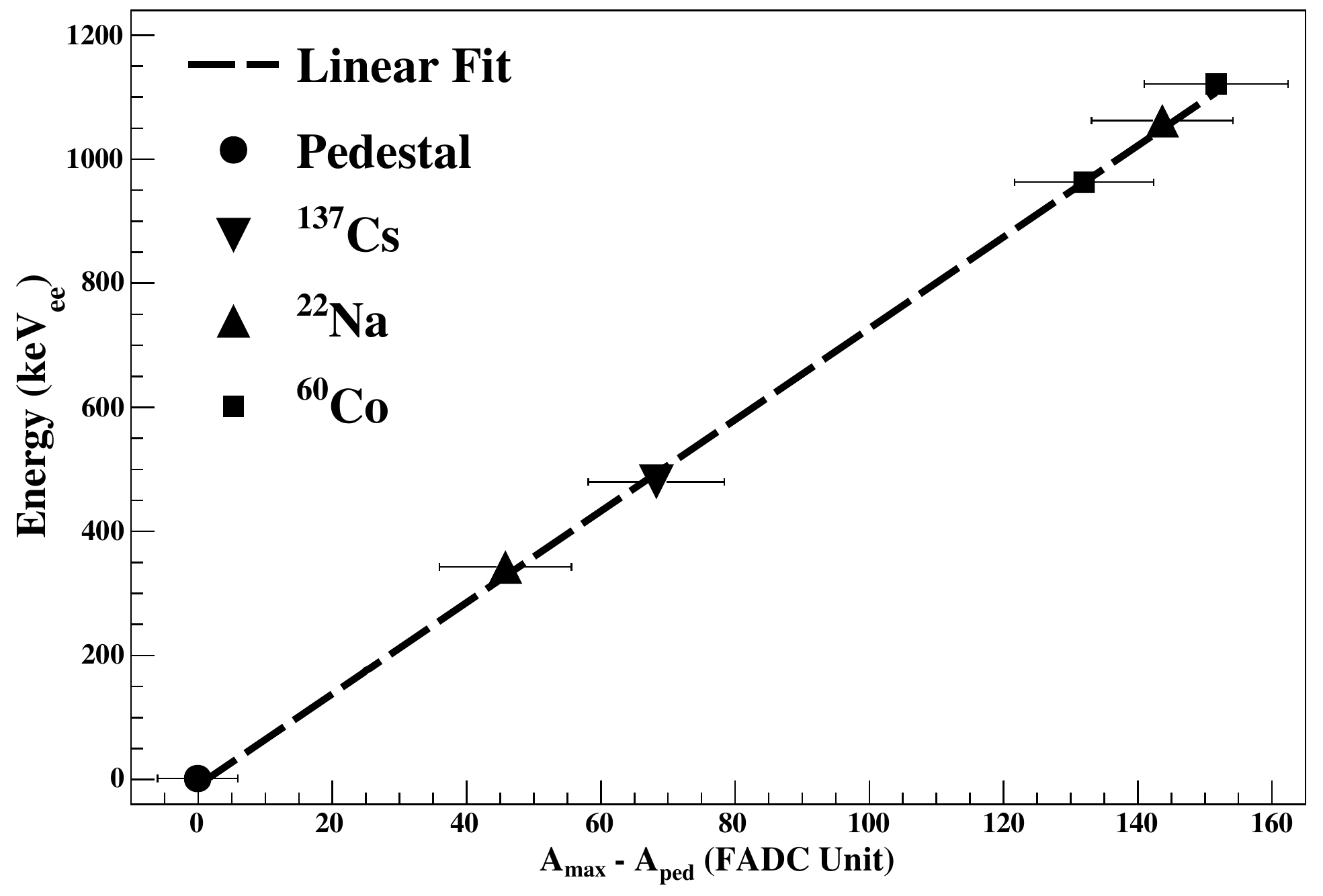}
\caption{Energy calibration of BC-501A liquid scintillator neutron detector in the parameter space of actual energy versus net amplitude that is after reduction of pedestal from the maximum amplitude.}
\label{fig::calib}
\end{figure}
\begin{figure}[hbt]
\includegraphics[width=8cm]{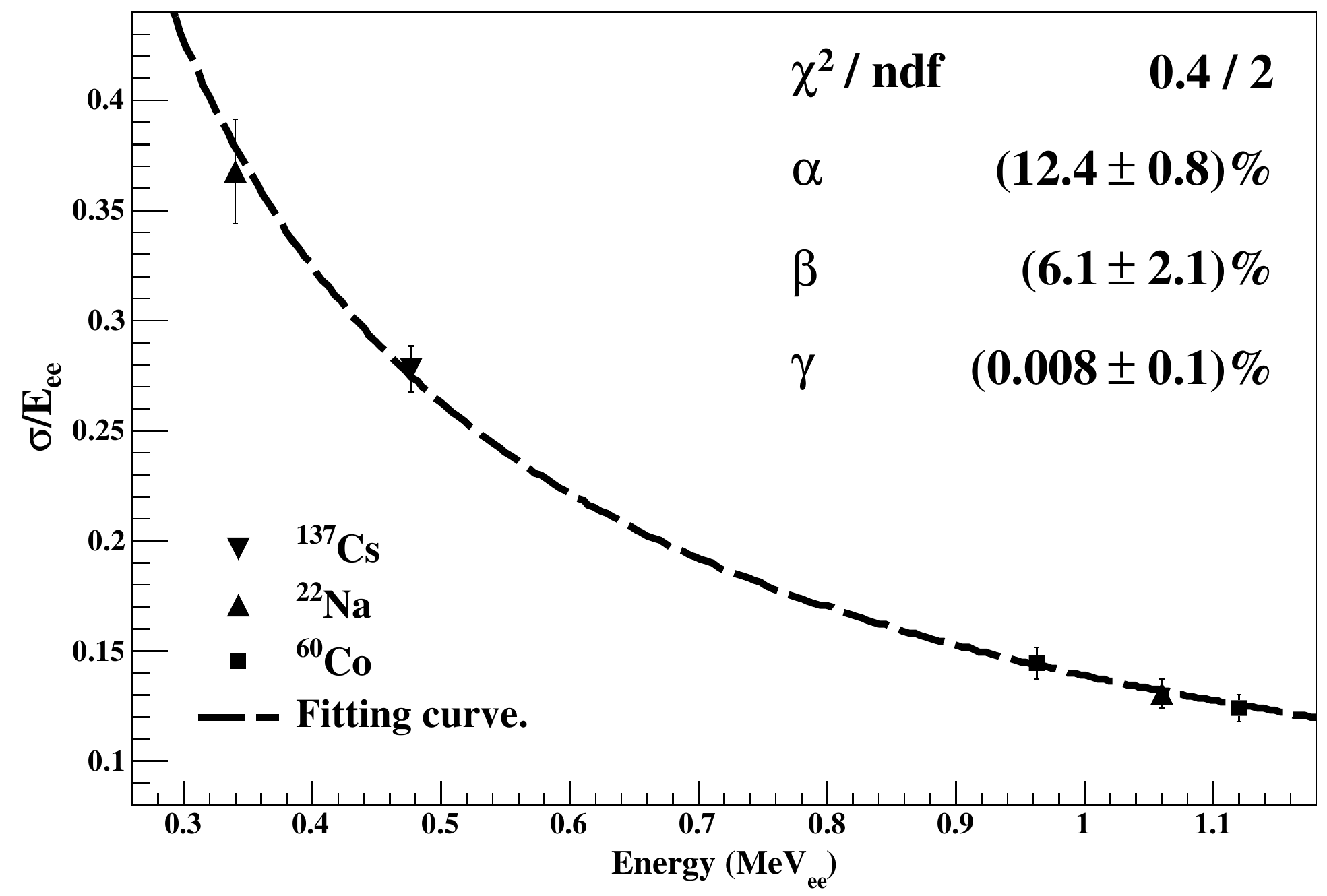}
\caption{Energy Resolution of the neutron detector.}
\label{fig::reso}
\end{figure}
\subsection{Energy Calibration}

The pulse height corresponds to the maximum voltage of the pulse which is linearly proportional to the energy of the incident particles. For the energy calibration of BC-501A liquid scintillator neutron detector, gamma ray from $^{22}Na$, $^{137}Cs$ and $^{60}Co$ sources are used. For the actual energy values of the sources their Compton edges energy values are accepted as listed in Table~\ref{tab::compt}. 

The calibration is performed based on the linear proportionality between the maximum pulse height and the energy of the events. Therefore, a first degree polynomial fit is performed between the net amplitude of the events, which can be defined as the remnant amplitude after substraction of pedestal from the pulse amplitude, and the actual energy values of the gamma rays. The linearity between actual energy values in electron equivalence (ee) unit and net amplitude in FADC unit accepted for the energy calibration procedure is illustrated in Figure~\ref{fig::calib}.

\begin{figure}
%\begin{center}
\textbf{(a)}\\
\includegraphics[width=8cm]{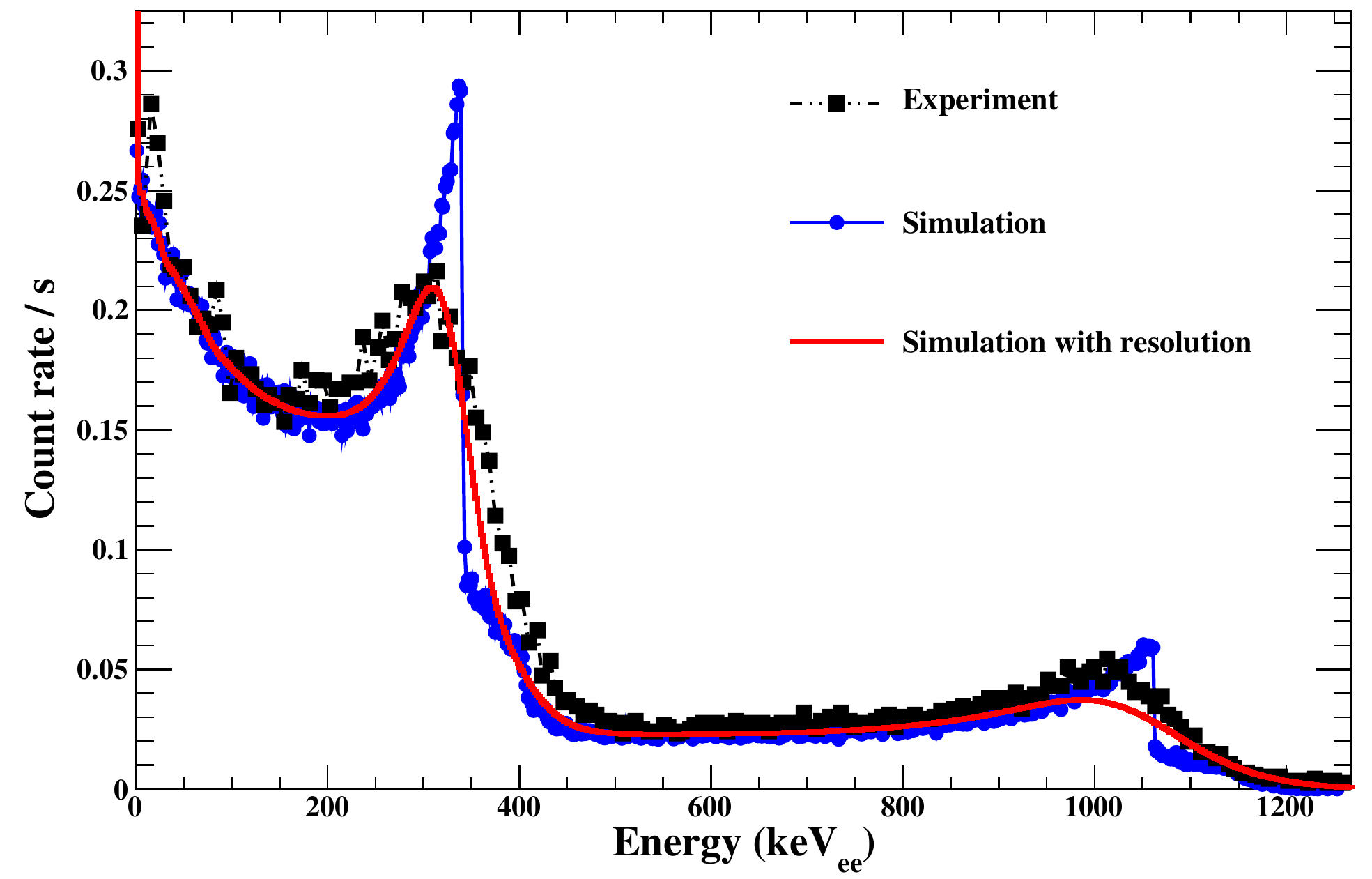}\\
\textbf{(b)}\\
\includegraphics[width=8cm]{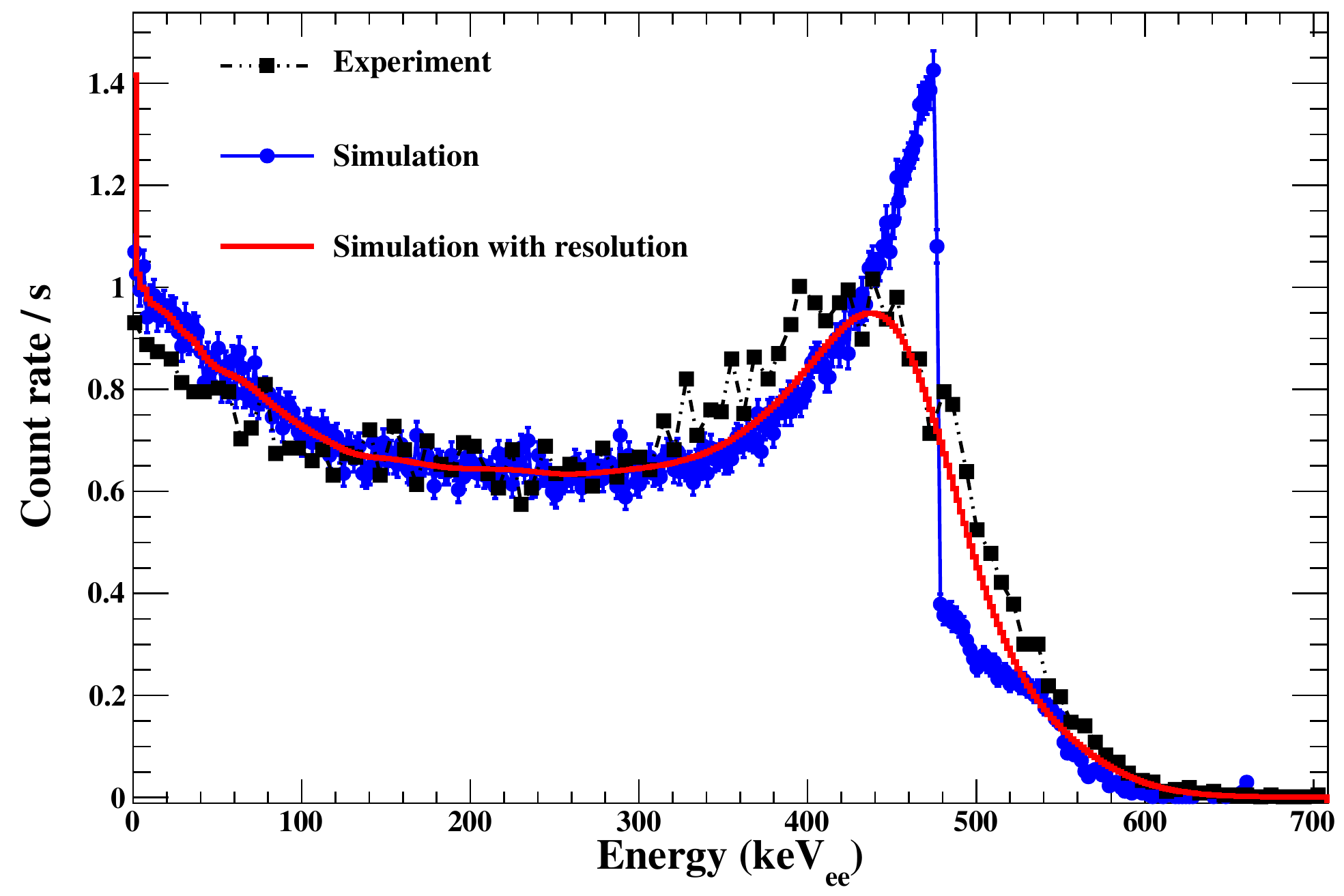}\\
\textbf{(c)}\\
\includegraphics[width=8cm]{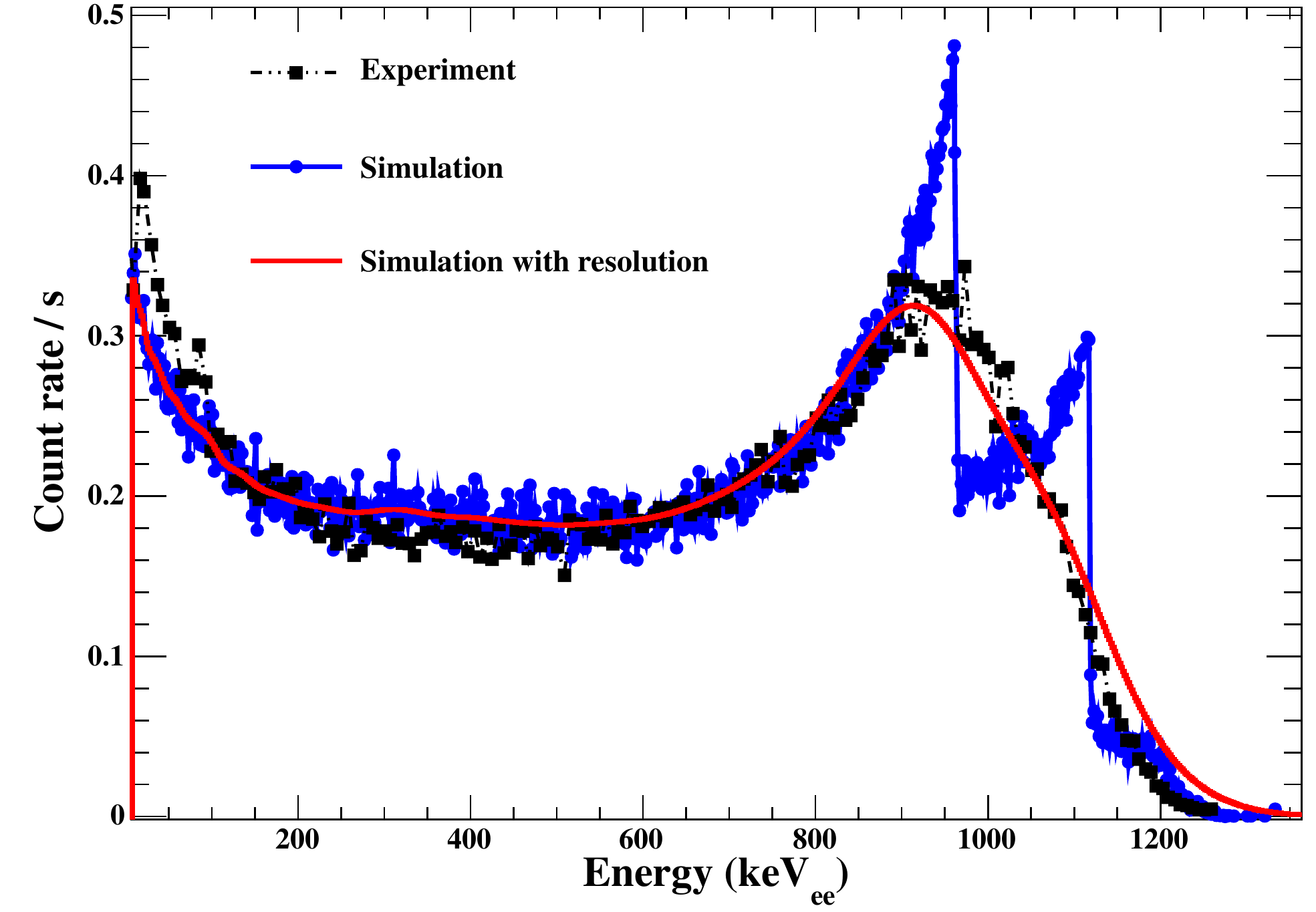}
\caption{(color online) Measured and simulated energy spectra without and with convoluted detector resolution for the source of (a) $^{22}Na$, (b) $^{137}Cs$, and (c) $^{60}Co$.}
\label{fig::spec}
%\end{center}
\end{figure}

The energy resolution is an important parameter to characterize any detector. The detector energy resolution in Full Width Half Maximum (FWHM) can be parameterized as,
\begin{equation}
\label{eq::res_fun}
\frac{dL(FWHM)}{L} = \sqrt{\alpha^{2} + \dfrac{\beta^{2}}{L} + \dfrac{\gamma^{2}}{L^{2}}},
\end{equation}
where $\alpha$ corresponds to the light transmission parameter depending on position in the detector, $\beta$ is related to the photoelectron production and $\gamma$ represents the electronic noise which is normally very small and can be ignored.

The measurement of the neutron detector energy resolution is illustrated in Figure~\ref{fig::reso}. The resolution parameters given in Eq.~\ref{eq::res_fun} are measured with the calibration sources. The measured value of the resolution parameters of the neutron detector are $\alpha = 12.4\%$, $\beta = 6.1\%$ and $\gamma = 0.008\%$. The calibrated energy spectra corresponding to different sources as well as the simulated energy spectra with and without detector resolution, are shown in Figure ~\ref{fig::spec}.
\begin{figure}[hbt]
\includegraphics[width=8cm]{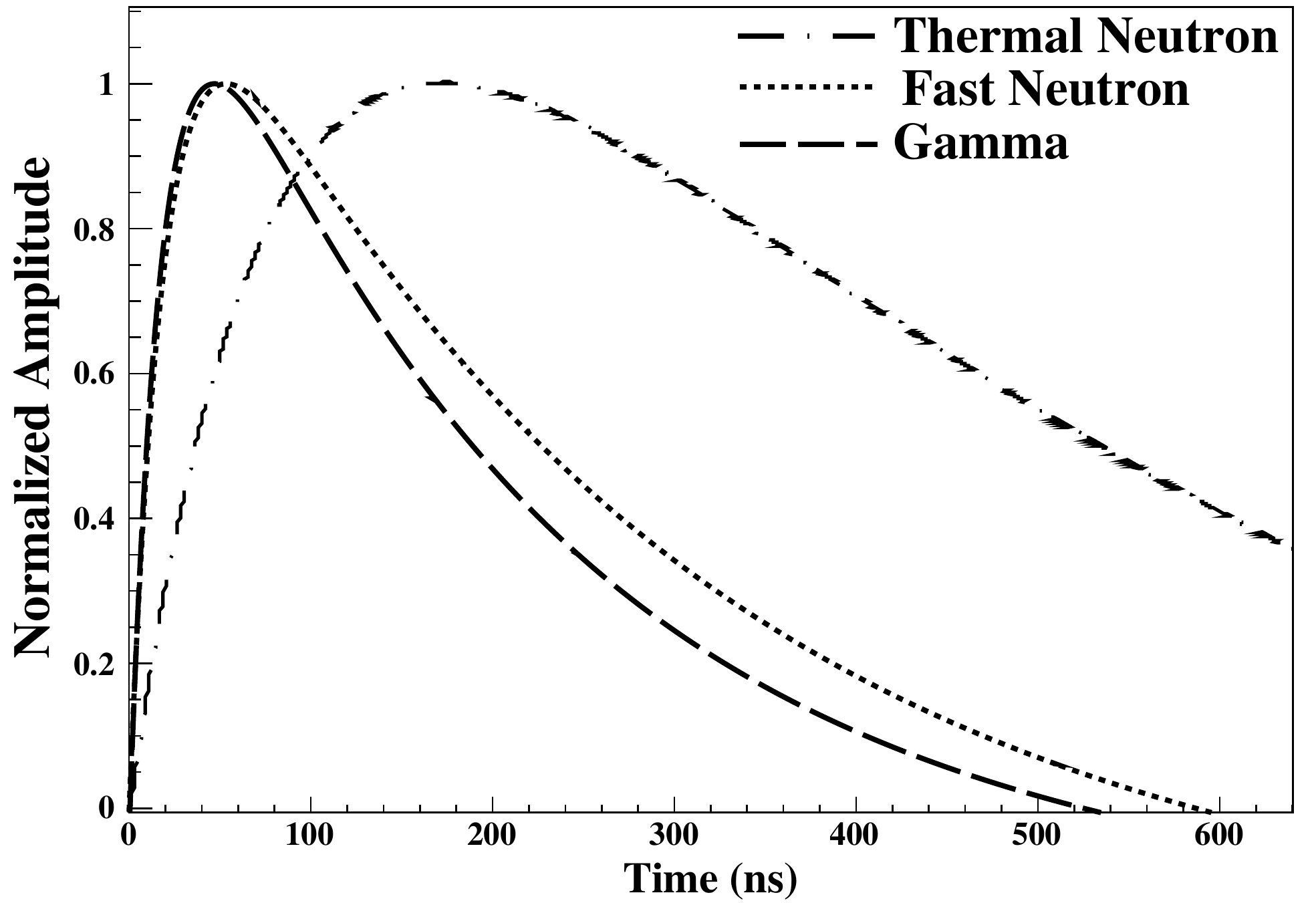}
\caption{Reference pulses for gamma, fast and slow neutron events from the inner to the outer, respectively.}
\label{fig::ref_pulse}
\end{figure}

\subsection{Neutron Response}

Three exponential function can explain the pulse shape and can be parameterized as~\cite{marrone},

\begin{eqnarray}
L &=& A\times(e^{-\theta(t-t_{0})}-e^{-\lambda_{s}(t-t_{0})}) \nonumber \\
&+& B\times(e^{-\theta(t-t_{0})}-e^{-\lambda_{l}(t-t_{0})}).
\label{eq::pshape}
\end{eqnarray}

Here L represents the pulse shape, A and B are the normalization constants, $t_{0}$ is reference time and $\theta$, $\lambda_{s}$, and $\lambda_{\ell}$ represent decay constants. $A/B$ ratio can describe different particles for a specific scintillator. Smoothed normalized reference pulse shapes representing different particles fitted by Eq.~\ref{eq::pshape} are shown in Figure~\ref{fig::ref_pulse}.

Owing to different energy deposition mechanisms of gamma rays, fast neutrons and thermal neutrons, the light creation timing produced by them would not be the same. Therefore discrimination between gamma rays, fast  neutrons and thermal neutrons can be obtained by examining the pulse shape information. By pulse shape analysis (i.e., rise and decay time of the pulses), we are able to categorize the events into three groups: gamma events, fast neutron events from BC501A and slow neutron events from BC702. Slow neutron events have the most characteristic signal output since ZnS(Ag) scintillation light output is dominated by a very slow decaying component for recoil $\alpha$ particles and other heavier ion contrary to the fast response of the organic scintillator detectors.  Fast neutron signals differ from that of $\gamma$ events slightly in their tail behavior, as seen in the Figure~\ref{fig::ref_pulse}.

\begin{figure}[hbt]
\includegraphics[width=8cm]{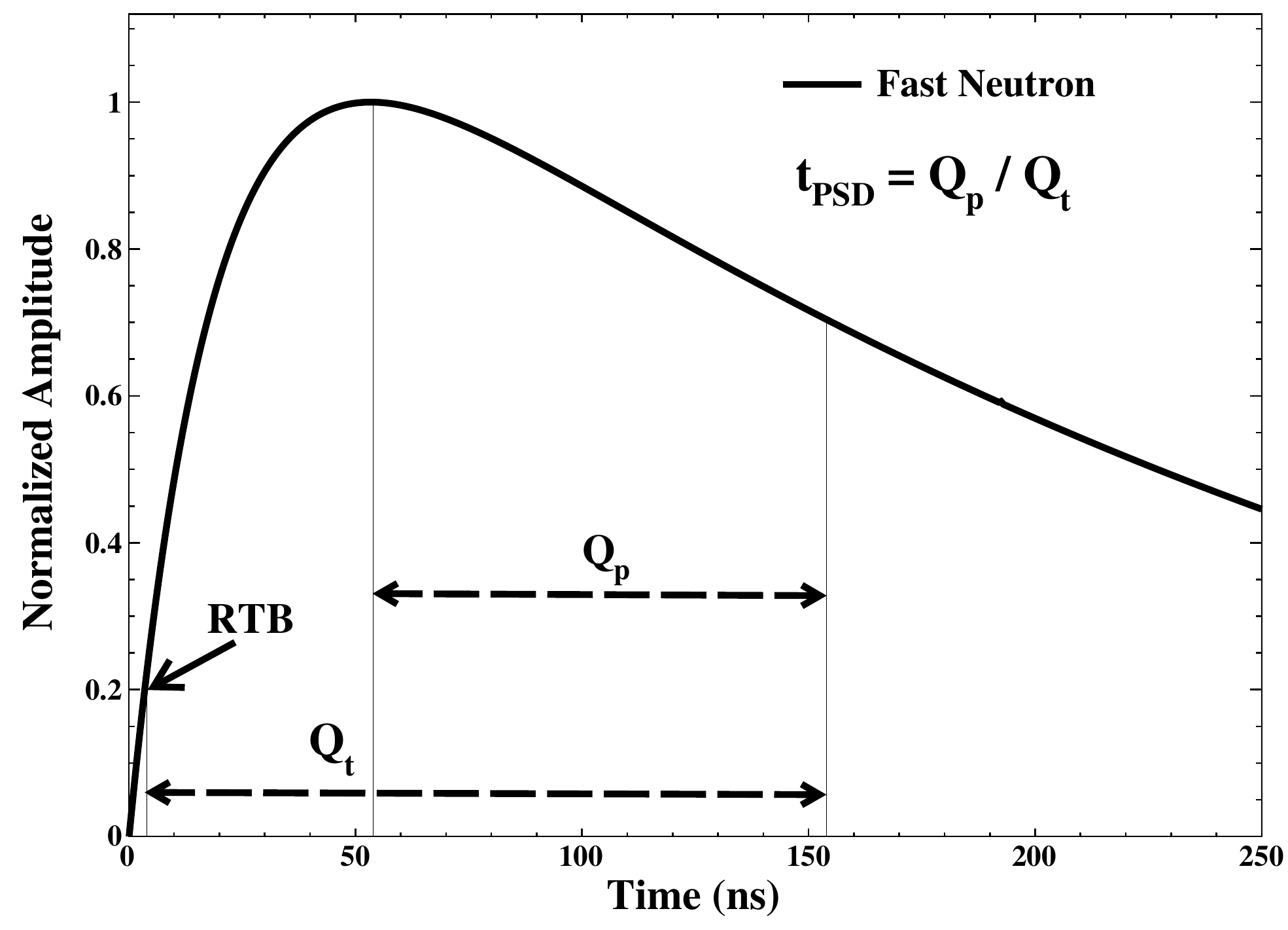}
\caption{Integral ranges for the calculation of PSD variable to differentiate gamma, fast and thermal neutron events. Vertical axis is in volts and normalized to unity. $Q_{p}$ is tail integration of the pulses and $Q_{t}$ is wider range integration of the pulses. The initial value of integration of each pulse depends on Rise Time Bin (RTB) that is defined where the pulses reach to 20\% of its own maximum value.}
\label{fig::lines}
\end{figure}
\begin{figure}[hbt]
%\begin{center}
{\bf (a)}\\
\includegraphics[width=8cm]{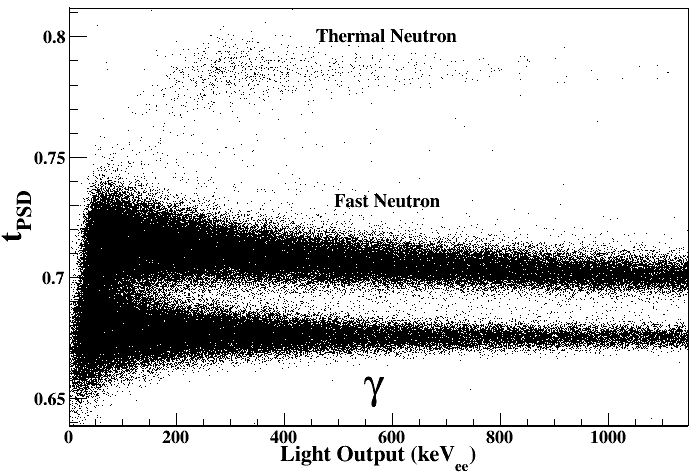} \\
{\bf (b)}\\
\includegraphics[width=8cm]{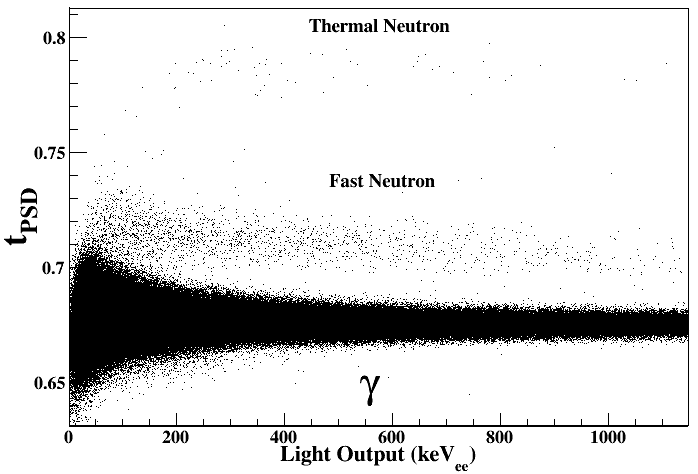}
\caption{Energy Distribution of the PSD variable for (a) $^{241}$AmBe($\alpha$,n) source, and (b) regular background. From top to bottom, three bands corresponding to three event groups, slow neutron, proton-recoil and electron recoil, respectively.}
\label{fig::psdDist}
%\end{center}
\end{figure}
\subsection{Event Identification}

In this study, a PSD technique is developed to identify the events orginated from different particle sources. In this PSD technique the pulses are integrated within two different time intervals and the ratio between them is taken. The best integration region is chosen where the distinction between the pulses are the most obvious. The initial point of the integration is the time-bin that the pulses reach to 20\% of their maximum amplitude and can be defined as rise time-bin (RTB). In order to differentiate observed events, two regions are selected for each individual pulses: first narrow range is the region where the distinction of the pulses is maximal and the other one has wider range starting from rising time bin (RTB) of the pulses. For both integration regions the initial point depends on RTB. Once the range of regions for the pulses are defined, the ratio of the two different partial integrals can be performed. By using the PSD technique the identification of the events can be maximized. Due to the fluctation of tail part of individual pulses especially for the pulses that correspond to the low energy, the partial integral range is chosen narrow in order to differentiate of gamma and neutron efficiently at low energies.

For the PSD study, a parameter named $t_{PSD}$, is used as given in Eq.~\ref{eq::psd}, which depends on RTB, since it yields better distinction between the pulses. The first integration region is defined as an integration region between 50 ns and 150 ns delay from the RTB respectively, which is denoted by $(Q_{p})$. The second integration region is defined as from RTB to 150 ns delay from RTB which is denoted by $(Q_{t})$. Therefore $t_{PSD}$ can be defined as simply taking the ratio between $Q_{p}$ and $Q_{t}$.

\begin{equation}
t_{PSD}=\frac{Q_{p}}{Q_{t}} = \frac{I[(RTB+50):(RTB+150)]}{I[(RTB) : (RTB+150)]}\label{eq::psd}.
\end{equation}

$^{241}$AmBe($\alpha$,n) is used as a reference source for gamma, fast neutrons and slow neutrons. Adopting the PSD technique given in Eq.~\ref{eq::psd}, we can see three spectral bands clearly appeared in Figure~\ref{fig::psdDist}. These three different bands corresponding to $\gamma$, fast neutron and slow neutron events from bottom to top, respectively.

The energy distributions of $^{241}$AmBe($\alpha$,n) and typical background are demonstrated in Figure~\ref{fig::psdDist}a and Figure~\ref{fig::psdDist}b, respectively. Well separated three bands of the reference $^{241}$AmBe($\alpha$,n) source and background data illustrates that the identification of gamma, fast neutron and thermal neutron events is successfully achieved.

\begin{figure}[hbt]
%\begin{center}
{\bf (a)}\\
\includegraphics[width=8cm]{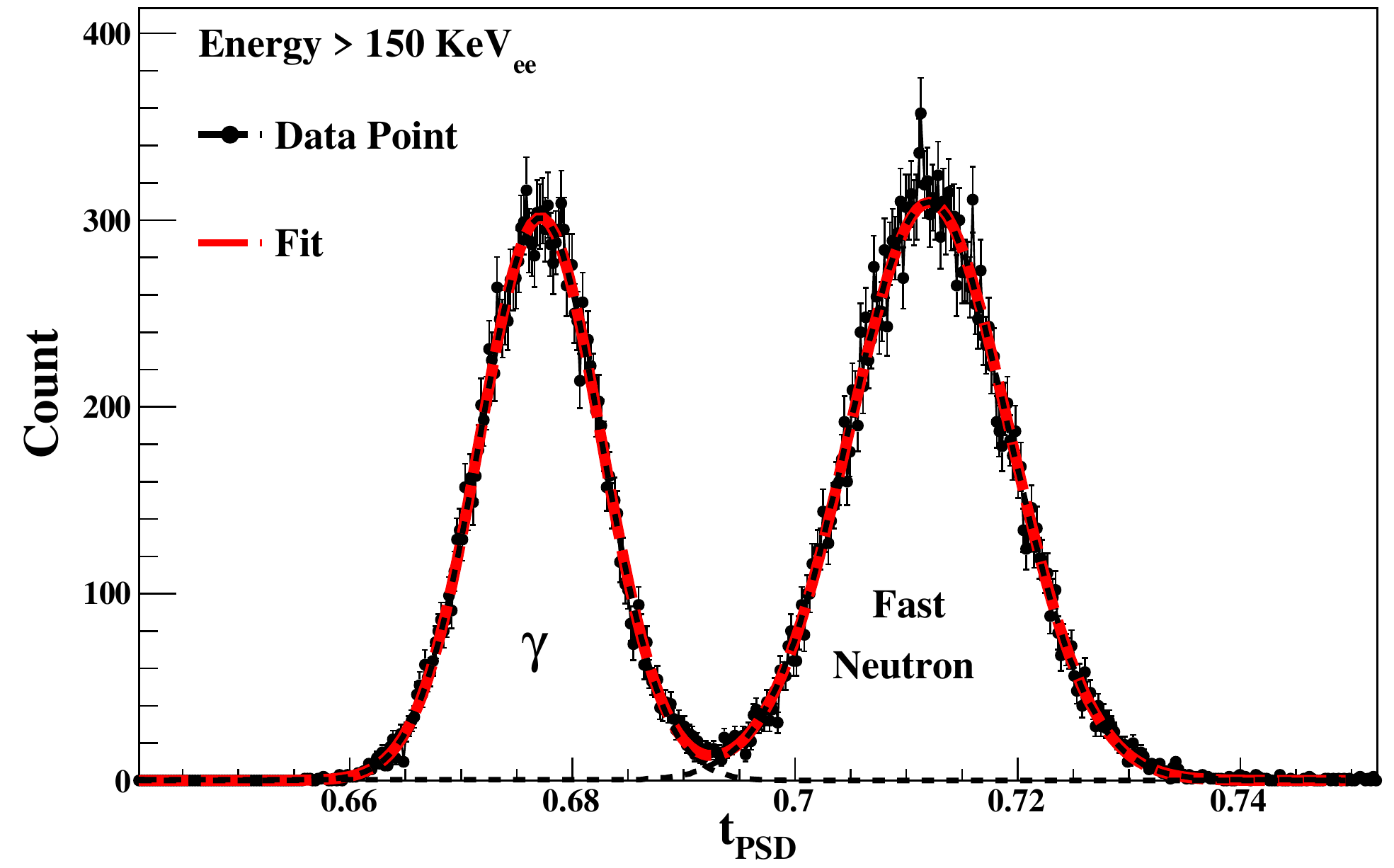} \\
{\bf (b)} \\
\includegraphics[width=8cm]{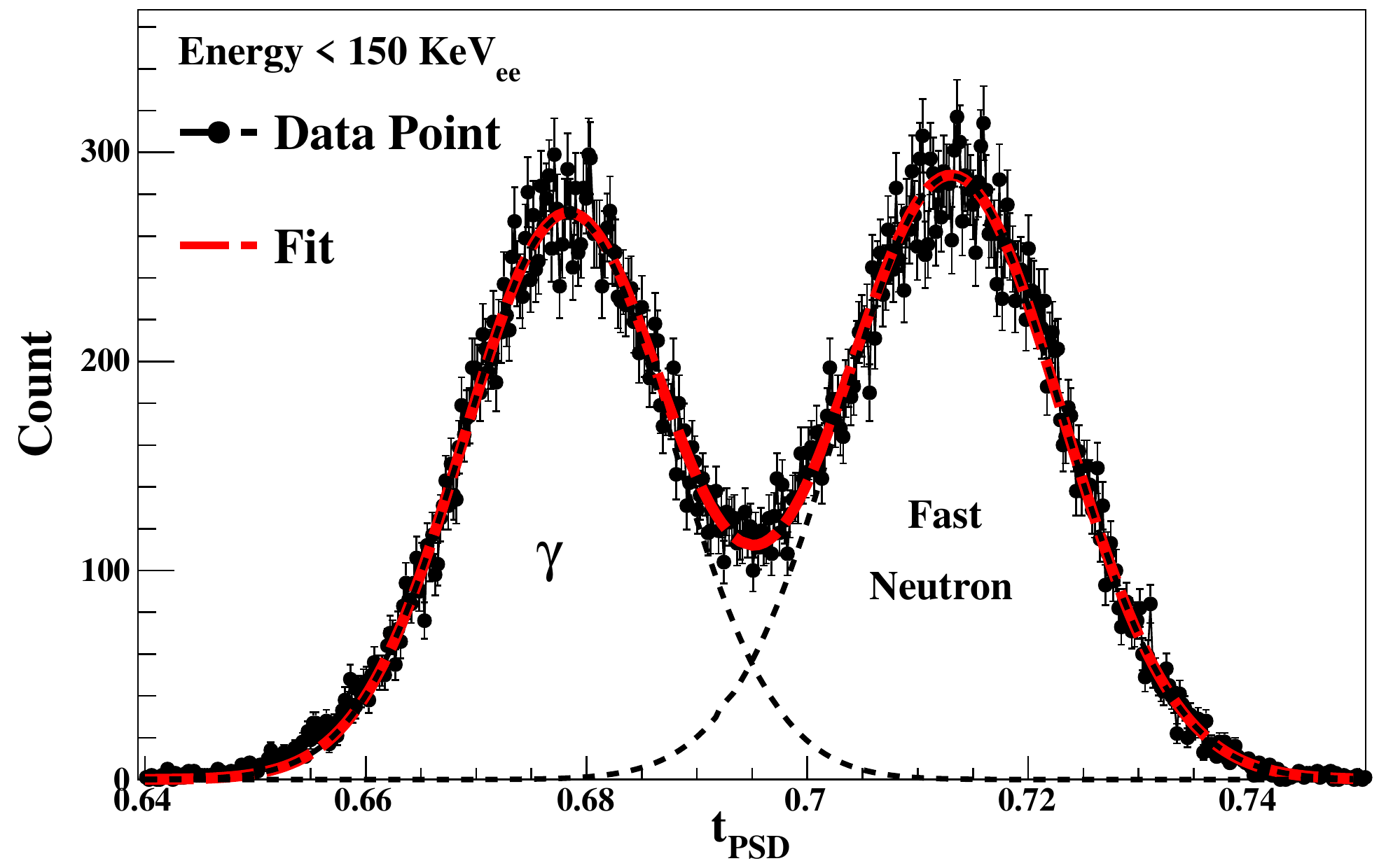}
\caption{\label{fig::psd_evnt} (color online) The PSD variable distributions for events of (a) $E>150$ $keV_{ee}$, and (b) $E<150$ $keV_{ee}$.}
%\end{center}
\end{figure}

Figure~\ref{fig::psdDist} reveals that the thermal neutron are well separated from fast neutrons as well as from gamma events, while fast neutron is well separated from gamma upto 150 $keV_{ee}$ and bellow 150 $keV_{ee}$ there are some leackage between fast neutrons and gamma events. The $t_{PSD}$ distributions of the events within different energy intervals are shown in Figure~\ref{fig::psd_evnt}, where Gaussian behaviors of both event groups are evident. Separation between fast neutrons and gamma events become very distinct above 150 $keV_{ee}$ energy range, as shown in Figure~\ref{fig::psd_evnt}a, in which a complete separation between the bands is possible. While below 150 $keV_{ee}$, as shown in Figure~\ref{fig::psd_evnt}b, a further separation for the combined Gaussian analysis is necessary.

In order to determine the quality of the separation Figure of Merit (FoM) as a parameter can be introduced as,
\begin{equation}\label{eq::fom}
FoM = \frac{(mean)_{n}-(mean)_{\gamma}}{(FWHM)_n+(FWHM)_{\gamma}},
\end{equation}
where the mean and FWHM can be obtained from the two Gaussian fit defined as,
\begin{equation}\label{eq::GG}
\frac{1}{\sqrt{2\pi\sigma_1^2}}e^{-\frac{(x-x_1)^2}{2\sigma_1^2}} + \frac{1}{\sqrt{2\pi\sigma_2^2}}e^{-\frac{(x-x_2)^2}{2\sigma_2^2}}.
\end{equation}

\begin{figure}[hbt]
%\begin{center}
\includegraphics[width=8cm]{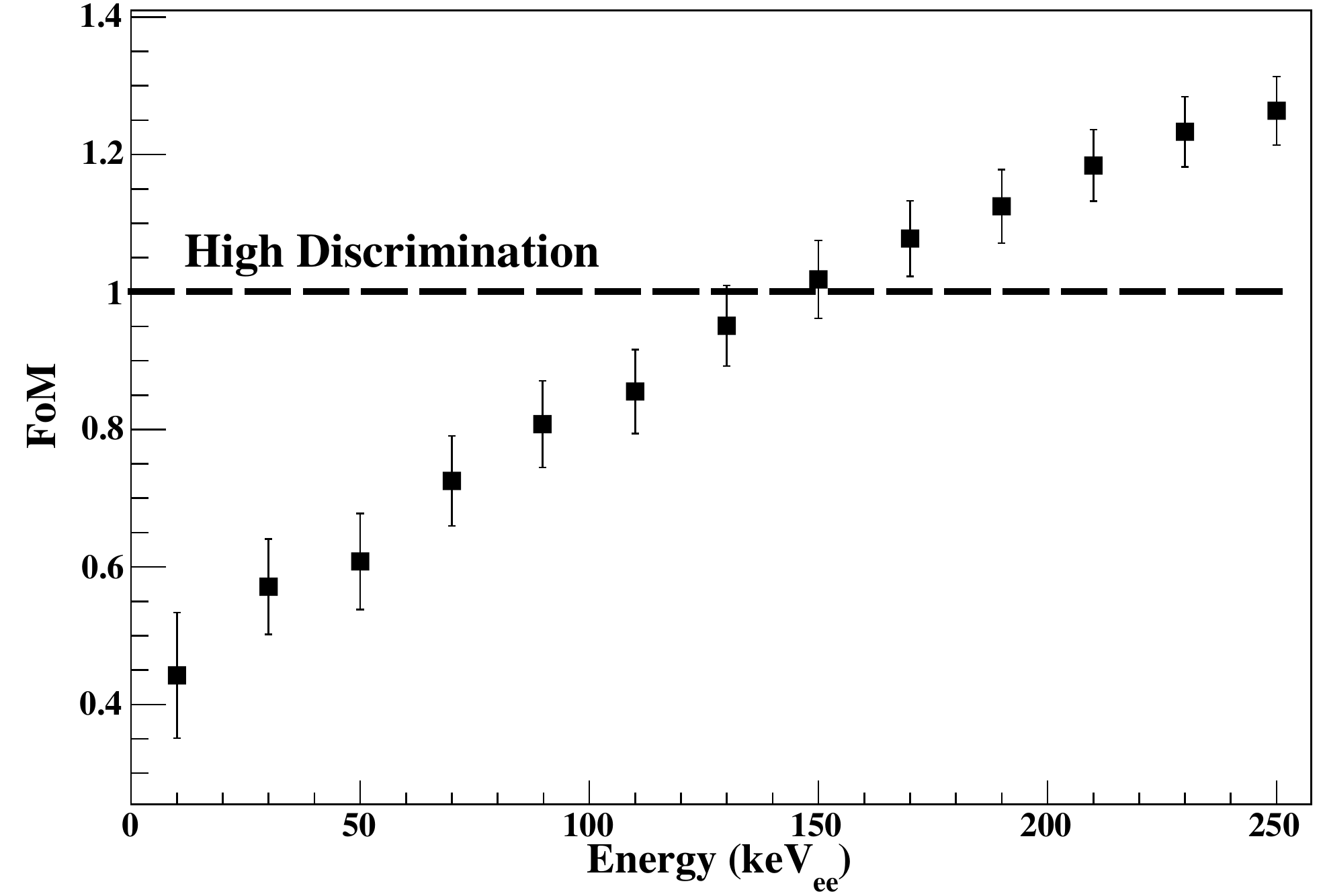}
\caption{The Figure of Merit (FoM) versus light output.}
\label{fig::fom}
%\end{center}
\end{figure}

The FoM given in Eq.~\ref{eq::fom} with respect to energy is illustrated in Figure~\ref{fig::fom}. FoM greater than unity is the indication of well separation. From Figure~\ref{fig::fom} it can be seen that above 150 $keV_{ee}$ the distinction is very clear and below 150 $keV_{ee}$ the two peaks of neutron and gamma start to merge.

\begin{figure}[hbt]
\begin{center}
{\bf (a)}\\
\includegraphics[width=8cm]{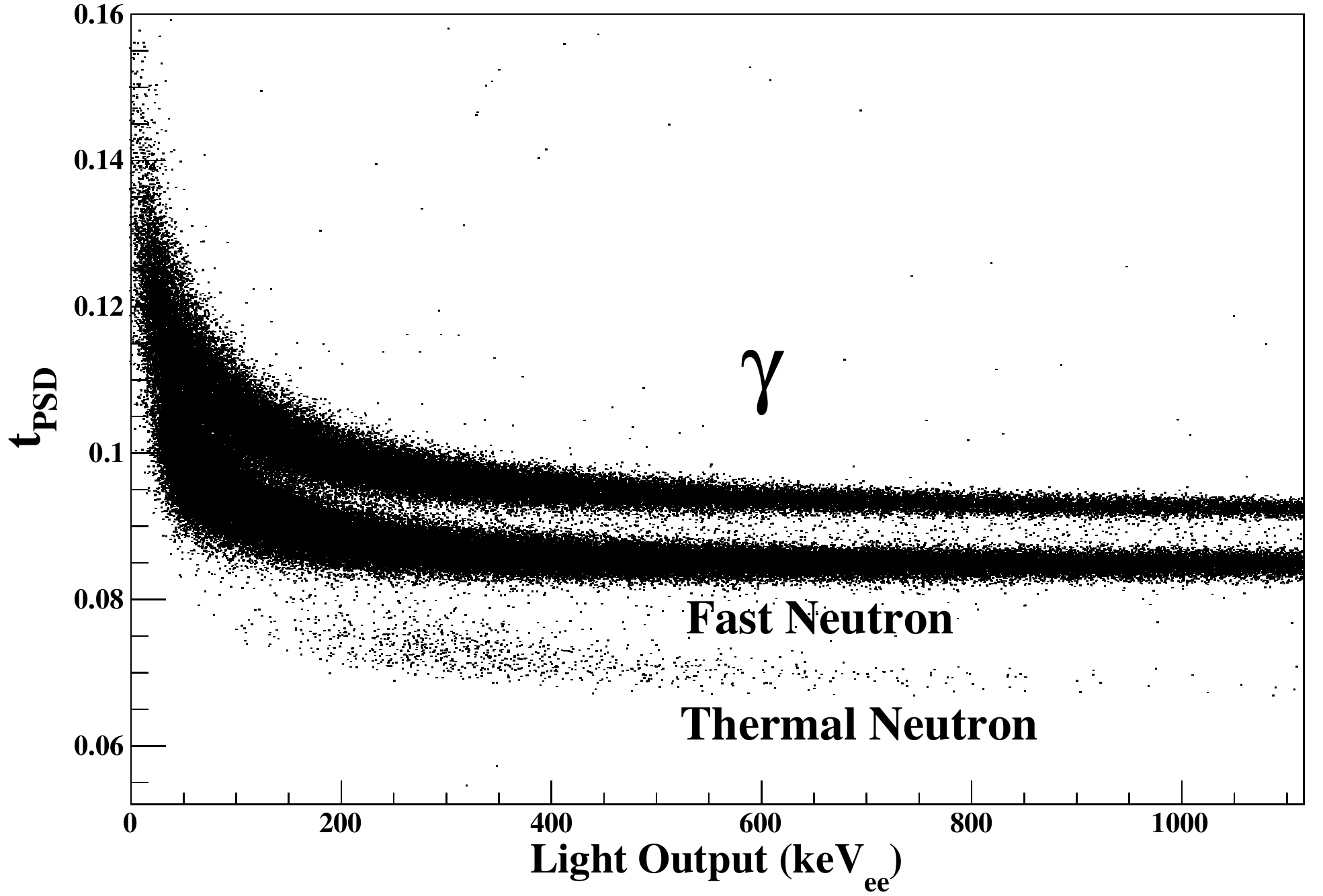} \\
{\bf (b)}\\
\includegraphics[width=8cm]{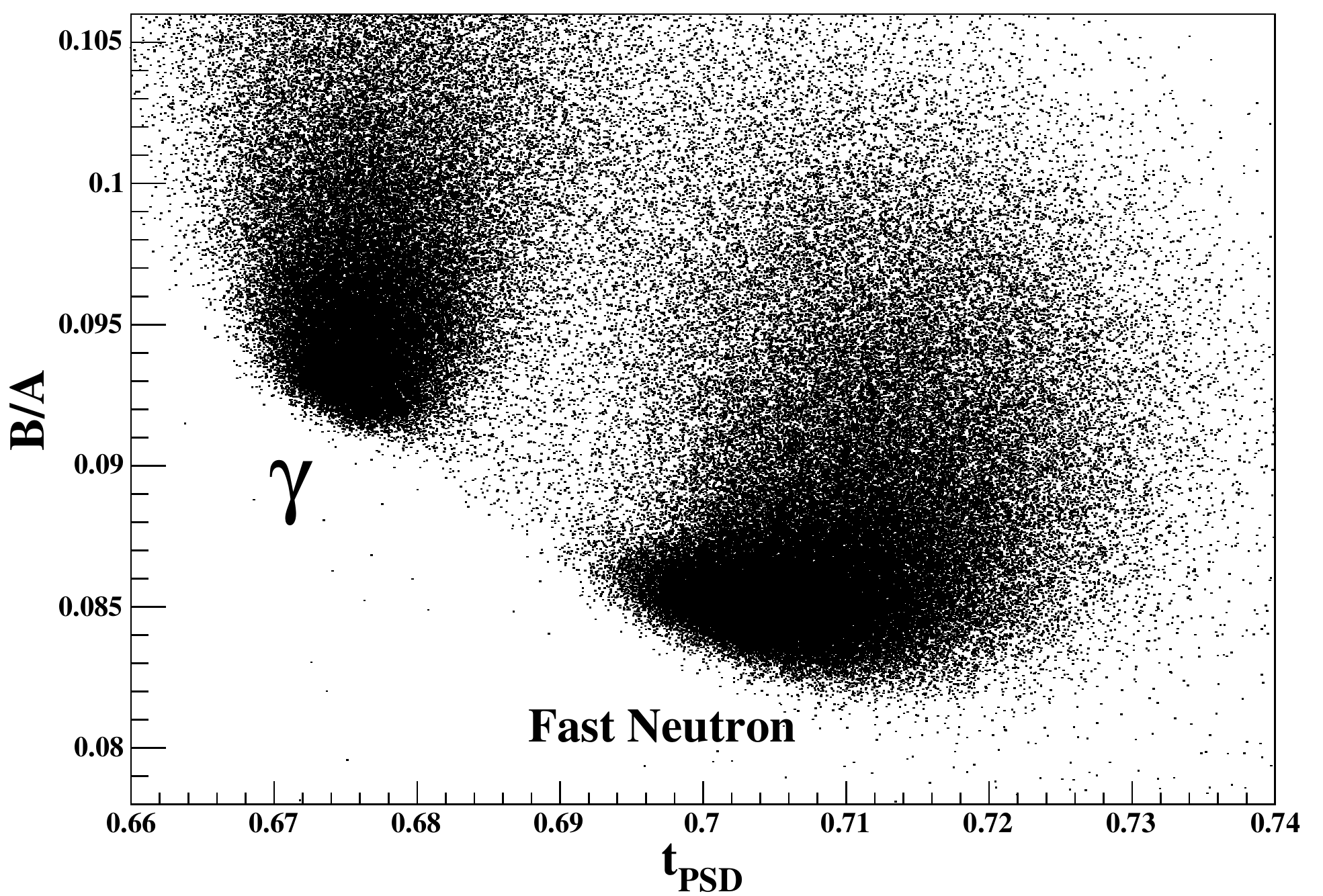} \\
\caption{The event distribution of $^{241}$AmBe source (a) in the parameter space of B/A with respect to energy, and (b) in the two different PSD method parameter spaces of $t_{PSD}$ and $B/A$.}
\label{fig::BA}
\end{center}
\end{figure}

Adopting the parameter of $L$ given in Eq.~\ref{eq::pshape} as a representative pulse shape, we can differentiate gamma events from the neutron events by working on the $B/A$ ratios of individual pulses accordingly. For this study a reference pulse is constructed by adopting $^{60}Co$ as the gamma source. The parameters of decay constants $\theta$, $\lambda_{s}$, $\lambda_{\ell}$ and reference time $t_{0}$ are obtained from the fitting of the gamma reference pulse. Then the pulse shape can be parametrized as,
\begin{eqnarray}
L &=& A\times\left[(e^{-(t-0.52)/226.6}-e^{-(t-0.52)/17.23}) \right.\nonumber \\
&+& \left.0.115(e^{-(t-0.52)/226.6}-1)\right],
\label{eq::BA}
\end{eqnarray}
where  $A$ is the only one free normalization parameter remains. The last exponential term in the Eq.~\ref{eq::pshape} goes to unity since $\lambda_l=3.72\times10^{-14}$. Individual pulses are fitted with function given in Eq.~\ref{eq::BA}. Corresponding PSD method for the identification of gamma against neutron events in $B/A$ parameter space with respect to energy is illustrated in Fig.~\ref{fig::BA}(a). As given in Eq.~\ref{eq::BA} gamma events are located in around the $B/A$ ratio band value of 0.115.

The distribution of $^{241}$AmBe events in the parameter space of two independent PSD methods, which are $t_{PSD}$ given in Eq.~\ref{eq::psd} and $B/A$ given in Eq.~\ref{eq::pshape}, is illustrated in Fig.~\ref{fig::BA}(b). The consistency of the values of the PSD parameter for gamma and neutron events shows both methods choose the same events.

\begin{figure}[hbt]
\begin{center}
\includegraphics[width=8cm]{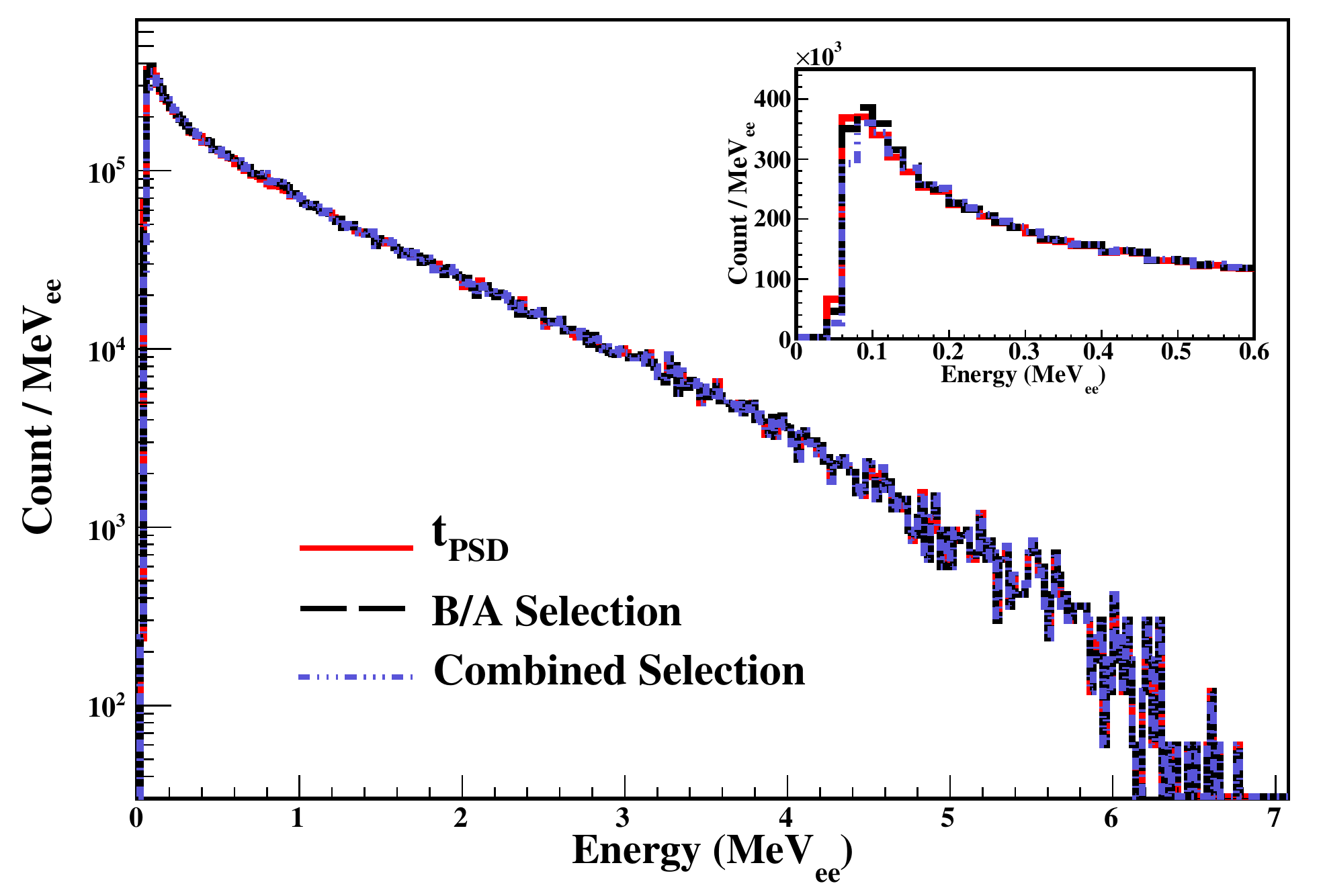}
\caption{The measured recoil spectrum of fast neutron events of $^{241}$AmBe source selected by different PSD methods. Combined selection corresponds to the selection in the two parameter space.}
\label{fig::psd_recoil}
\end{center}
\end{figure}

The measured neutron recoil spectra of $^{241}$AmBe based on the two independent PSD methods are illustrated in Fig.~\ref{fig::psd_recoil}. We also performed a selection of the events in the two parameter space of $t_{PSD}$ and $B/A$ shown in Fig.~\ref{fig::BA}(b) as a combined selection. As it can be seen that there is a perfect match in the recoiled spectra based on different selection criteria showing that both PSD methods choose the same fast neutron events.

\section{Unfolding and Reconstruction of Neutron Spectra}

In the measurement of neutron, detection carried under an arbitrary particle flux of $\Phi(E)$, the measured spectrum $N(L)$ can be obtained as,
\begin{equation}
N(L)= \int_{0}^{\infty} \Phi(E) \ R(L,E) \ dE,
\label{eq::meas_spec}
\end{equation}
where $E$ is the incident particle energy,  $L$ is the measured recoil energy and $R(L,E)$ is called the response function which correlates the flux and the measured spectrum. The response function for a fixed E can also be considered as the measured spectrum of a monochromatic beam of incident particles at energy $E$, or the probability distribution of the measured recoil energy for a single incident particle with kinetic energy $E$. Even if the form of $R(L,E)$ is known exactly for a detector system, the integral given in Eq.~\ref{eq::meas_spec} may not be solvable for $\Phi(E)$ analytically. For a real measurement, the equation should be in discrete form such as bellow,
\begin{equation}
N_{i}(L_{i})=\sum_{j} \Phi_{j}(E_{j}) \ R_{ij}(L_{i},E_{j}).
\label{eqn::rateD}
\end{equation}

Finding $\Phi_{j}(E_{j})$ requires developing a computational method called unfolding (or deconvolution) which is widely used in neutron spectroscopy and dosimetry applications. Response function can be constructed from the measurement of recoil spectra of monochromatic neutrons at every beam energy $E_{j}$. In this study, response function is calculated via Monte Carlo simulation giving light output and resolution functions as input parameters.
\begin{figure}[hbt]
%\begin{center}
\includegraphics[width=8cm]{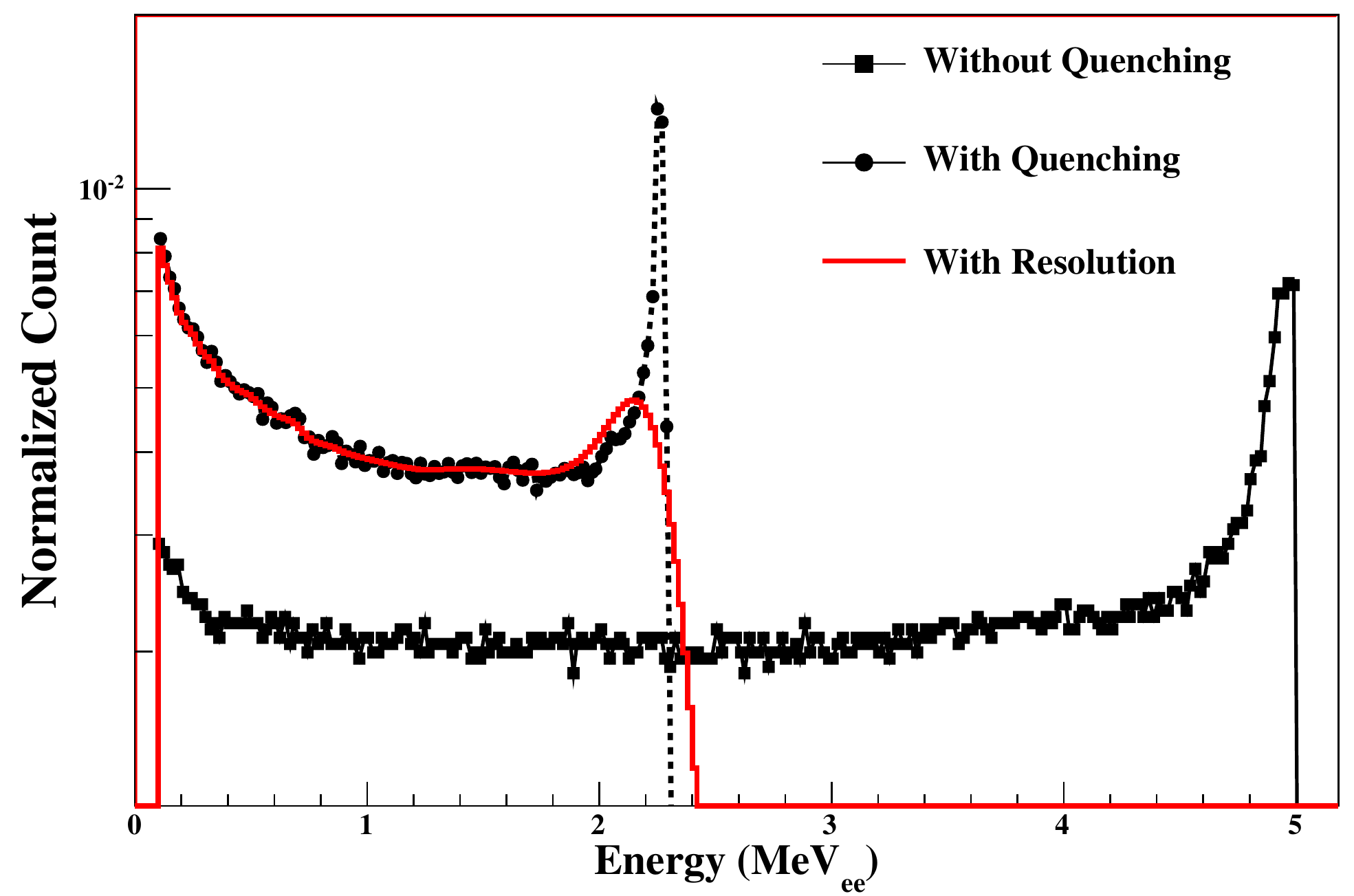}
\caption{(color online) Simulation of the detector response to 5 $MeV$ monochromatic neutron beam with and without quenching effect. Quenched recoiled  energy spectra after including detector resolution that is given in Eq.4 and shown in Fig.8  is also superimposed.} 
\label{fig::respFold}
%\end{center}
\end{figure}
\begin{figure}[hbt]
{\bf (a)}\\
\includegraphics[width=8cm]{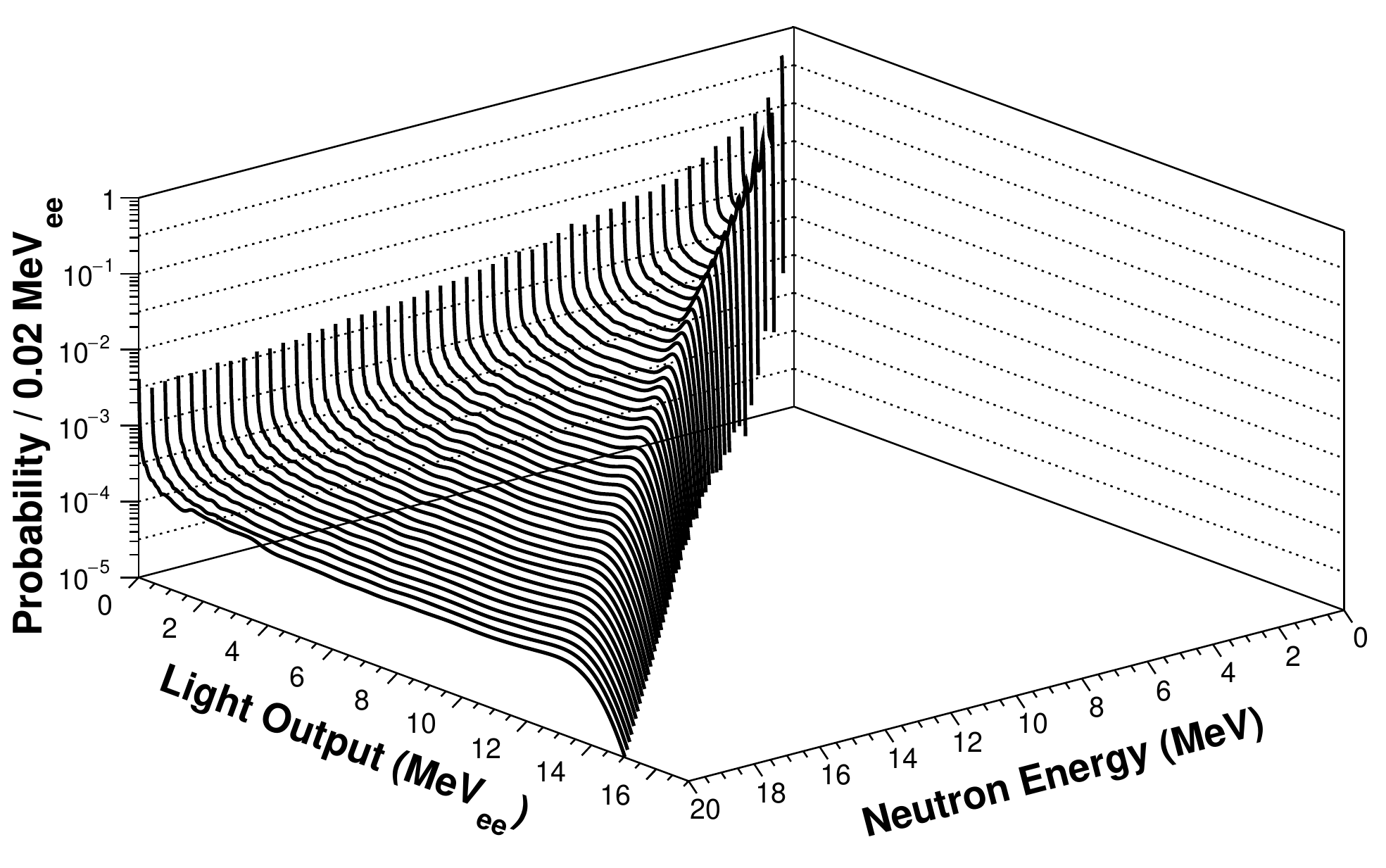}\\
{\bf (b)}\\
\includegraphics[width=8cm]{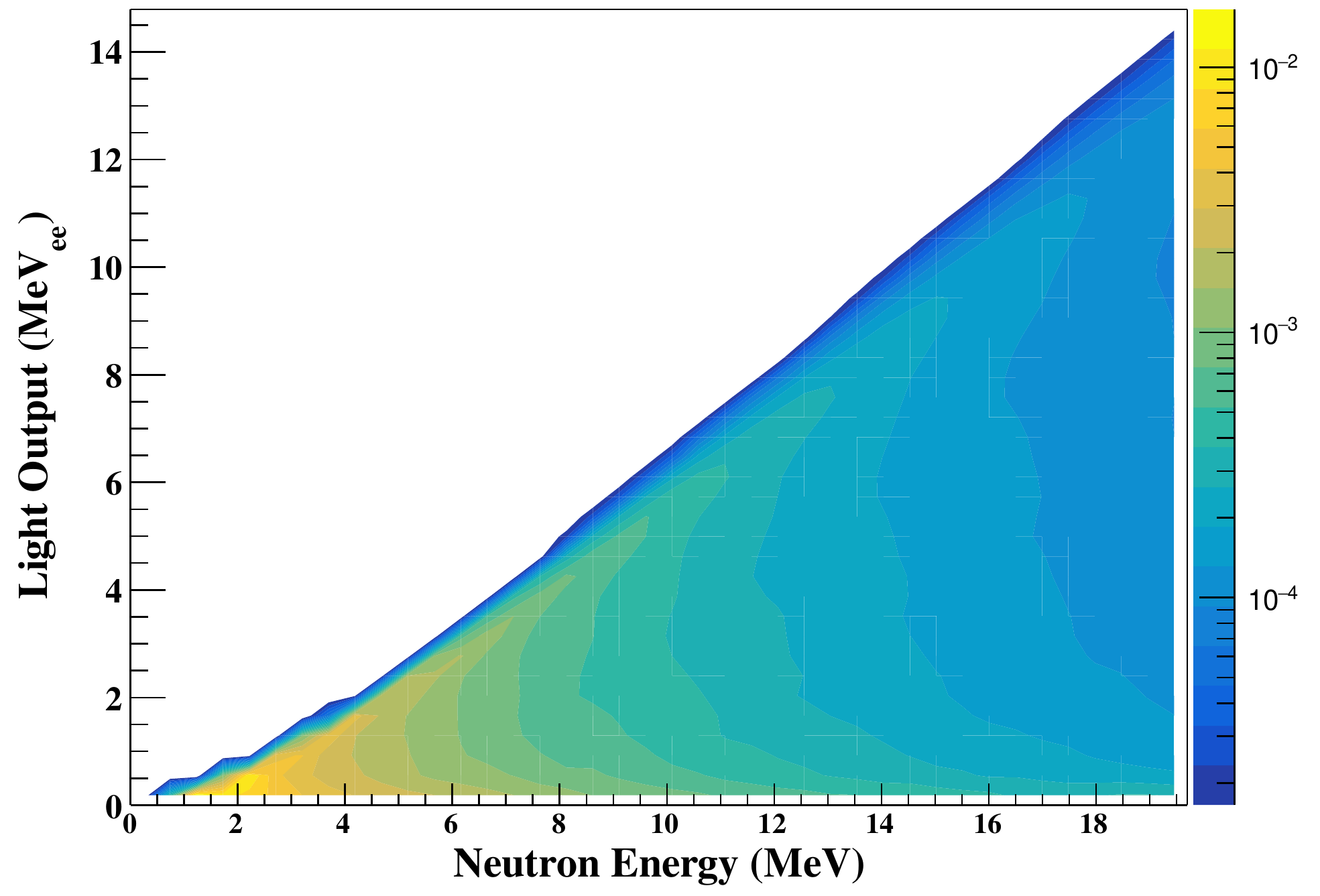}
%\end{minipage}
\caption{(color online) Response functions $R(L,E)$ for $E_n=0-20~MeV$ monochromatic neutrons (a) regular 2D, and (b) contour 2D modes.}
\label{fig::resp_all_neutron}
%\end{center}
\end{figure}
\begin{figure}[hbt]
{\bf (a)}\\
\includegraphics[width=8cm]{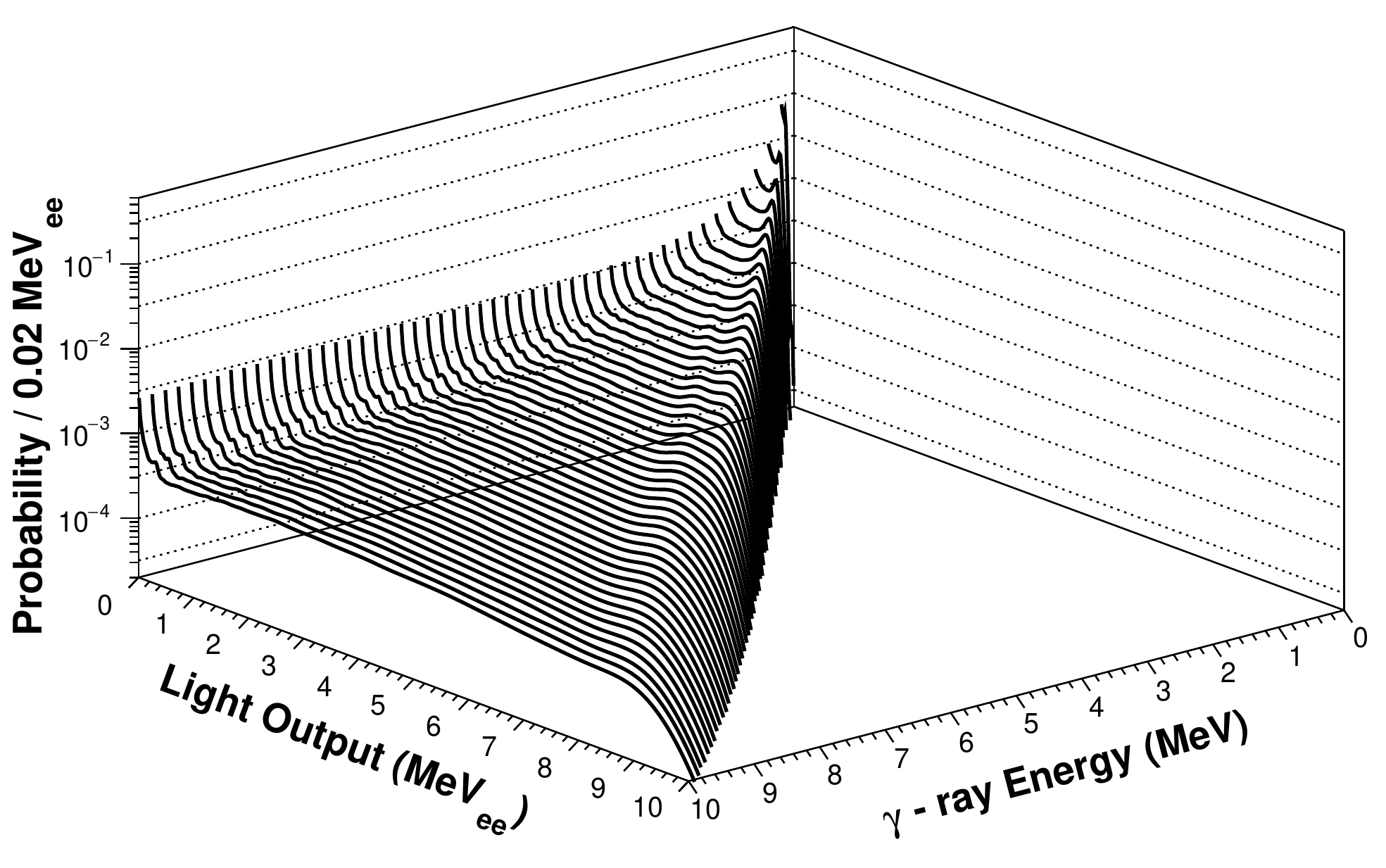}\\
{\bf (b)}\\
\includegraphics[width=8cm]{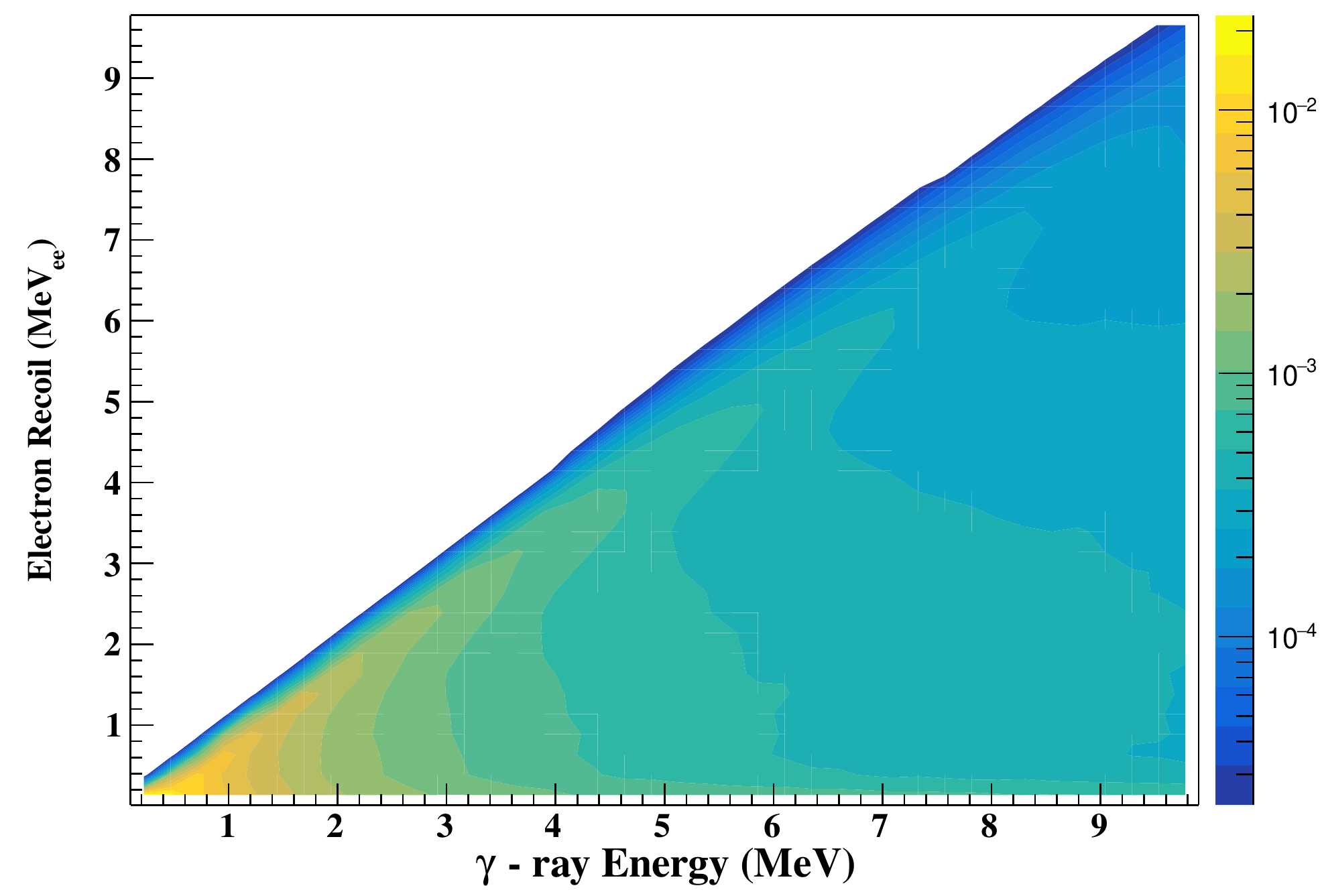}
%\end{minipage}
\caption{(color online) Response functions $R(L,E)$ for at $E_{\gamma}=0-10~MeV$ monochromatic gammas in (a) regular 2D, and (b) contour 2D modes.}
\label{fig::resp_all_gamma}
%\end{center}
\end{figure}
\subsection{Response Function and Unfolding}

Response function $R_{ij}(L_i,E_j)$ is generated by Monte Carlo simulation, by means of generating 1 million neutron particles at energy $E_j$ to obtain the light output spectrum L of recoil particles. Energy deposition by carbon nuclei should be ignored since they are highly quenched.

For the calculation of the neutron flux $\Phi(E_{j})$, the response functions are calculated for 1000 neutron beam energies of $E_{j}$ between 0 and 20 $MeV$. Indeed, the response to the neutrons starts from 100 $keV_{ee}$. For a specific neutron energy of incidence $E_{j}$, and $i^{th}$ component of light output $L_i$, the response function can be written as,
\begin{equation}
R_{ij}(L_{i},E_{j})=\sum_{k}\int_{L_{i}-\Delta L}^{L_{i}+\Delta L}\frac{C_{k}}{\sqrt{2\pi} \sigma}e^{\frac{-(L'-L_{k})^{2}}{2\sigma^{2}}}dL' \ ,
\label{eq::rf}
\end{equation}
where $L_i$ is the $i^{th}$ bin of light output value of the response function, the content of which is to be calculated. Contribution from each Gaussian distributed bin of the spectrum at $L_k$ is integrated in the interval of $L_i\pm \bigtriangleup L$. $C_k$ is the content of the $k^{th}$ bin of the spectrum. $\sigma(L_k)$ is the RMS error for $L_k$ which can be calculated from the resolution function given in Eq.~\ref{eq::res_fun} as,
\begin{equation}
\sigma(L)=\frac{dL(FWHM)}{2\sqrt{2~ln~2}}.
\end{equation}

As an illustration, Figure~\ref{fig::respFold} shows the response function for incident neutrons with 5 $MeV$ which is obtained from the corresponding simulated light output of recoil spectrum and the resolution functions. The response functions of R(L,E) are constructed for 1000 neutrons energies up to 20~$MeV_{ee}$ and for 500 gammas energies up to 10~$MeV_{ee}$ with 20~$keV_{ee}$ energy steps. The response functions for all neutron and gamma energies are illustrated in Figure~\ref{fig::resp_all_neutron} and Figure~\ref{fig::resp_all_gamma}, respectively. 

Unfolding is a widely used computational method in spectroscopy. The method had been developed for neutron / gamma dosimetry measurements for radiation protection purposes at nuclear facilities or similar environments where it is of crucial importance to know the abundance of the neutron / gamma radiation. There are several computational algorithms for unfolding in literature. In this work two of them are used, one is developed by Doroshenko ~\cite{doro} and the other one is Gravel unfolding method~\cite{gravel}, both of them calculate the neutron flux by an iterative method. The iteration algorithm of Doroshenko can be written as,
\begin{equation}
\Phi_{j}^{n+1}=\frac{\Phi_{j}^{n}}{\sum_{i}R_{ij}}\sum_{i} R_{ij} \frac{N_{i}}{\sum_{k} \Phi_{k}^{n}R_{ik} }.
\end{equation}
\begin{figure}[hbt]
{\bf (a)}\\
\includegraphics[width=8cm]{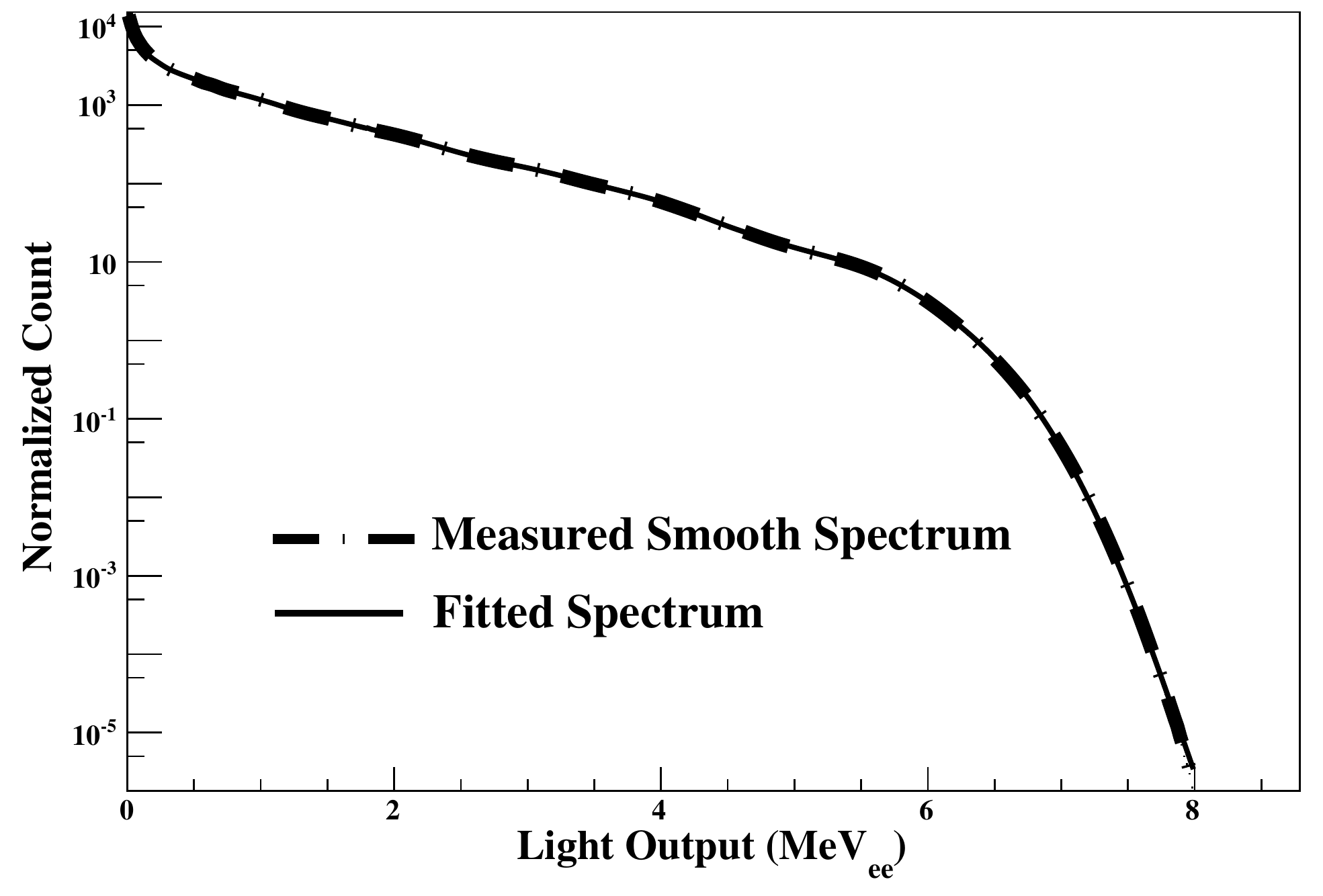} \\
{\bf (b)}\\
\includegraphics[width=8cm]{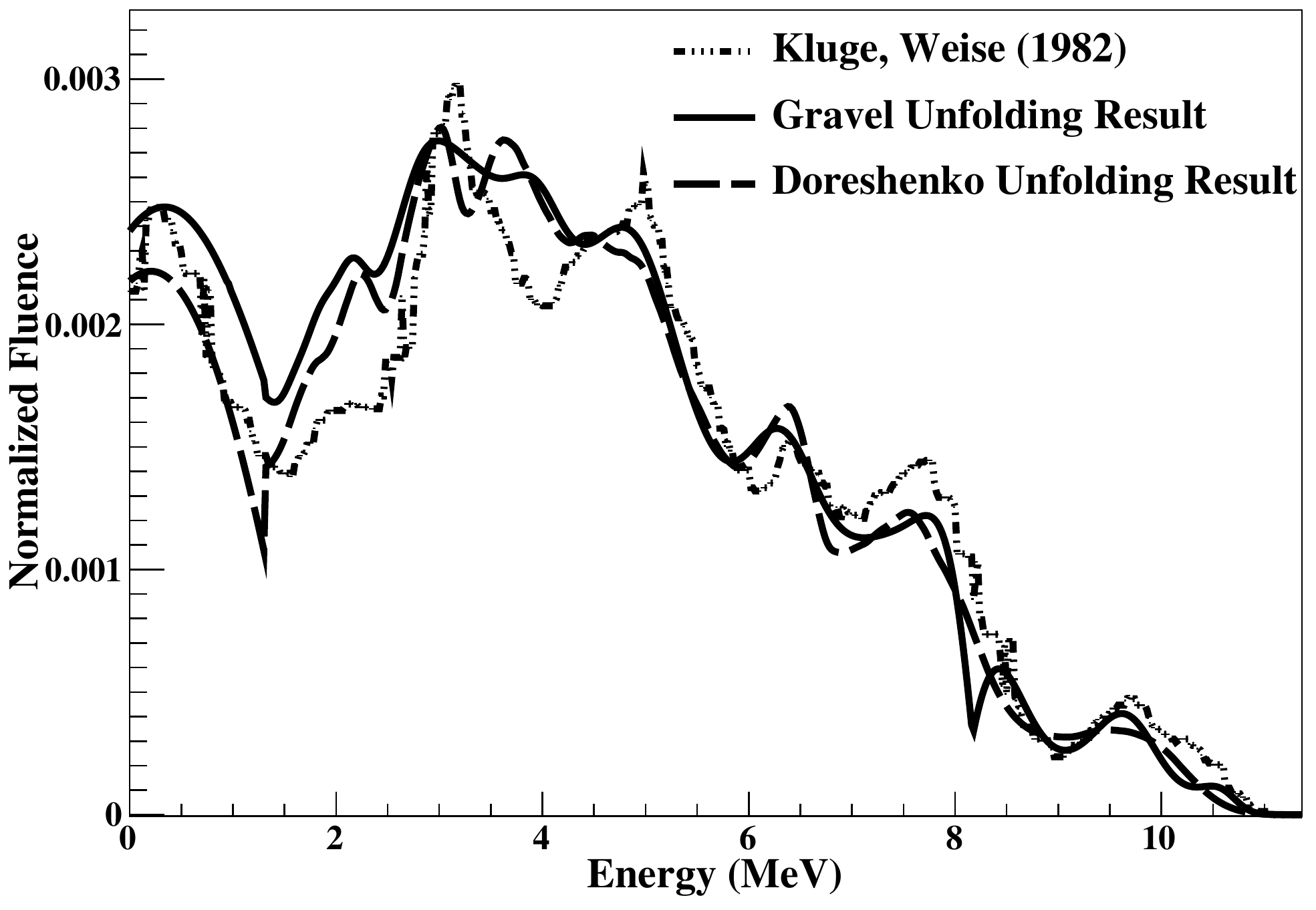}
%\end{minipage}
\caption{(a) Measured recoil spectrum of $^{241}$AmBe($\alpha$,n) source, (b) Convoluted neutron fluence of $^{241}$AmBe($\alpha$,n) source with two unfolding methods in comparison with the measurement of Kluge and Weise (1982)~\cite{kluge82}. The two spectra are normalized to equal number of detector counts.}
\label{fig::ambe}
%\end{center}
\end{figure}
\begin{figure}
{\bf (a)}\\
\includegraphics[width=8cm]{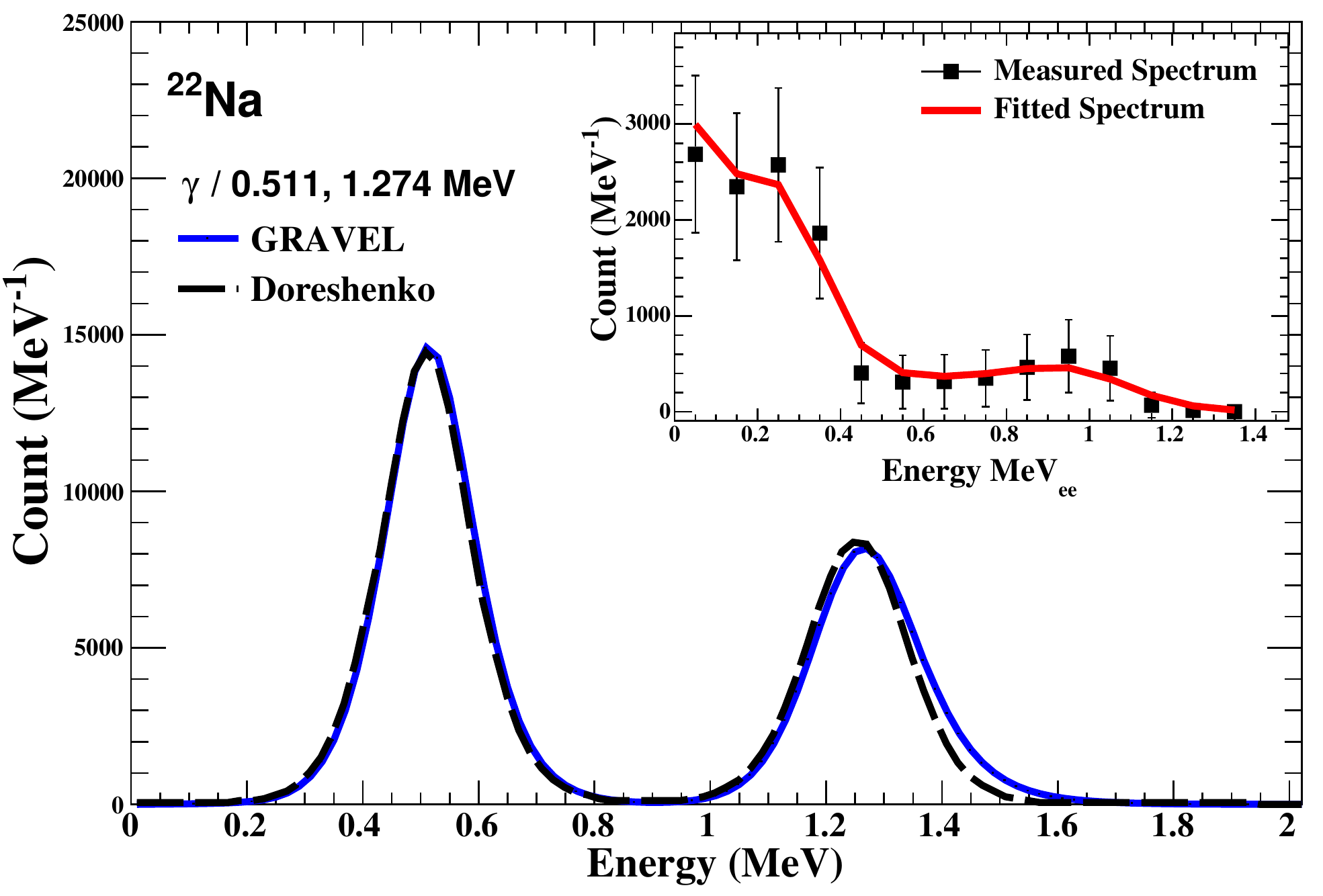} \\
{\bf (b)}\\
\includegraphics[width=8cm]{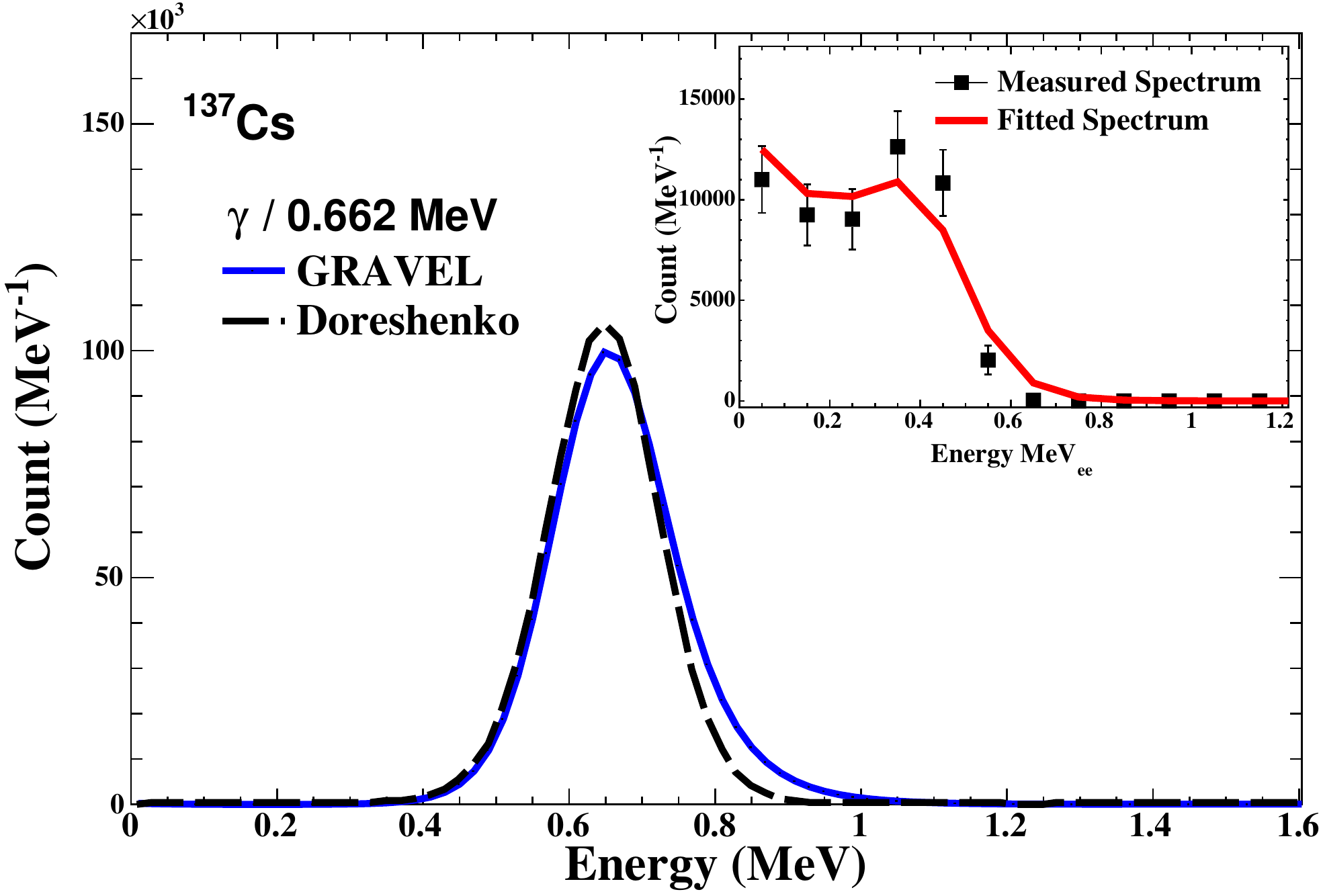}\\
{\bf (c)}\\
\includegraphics[width=8cm]{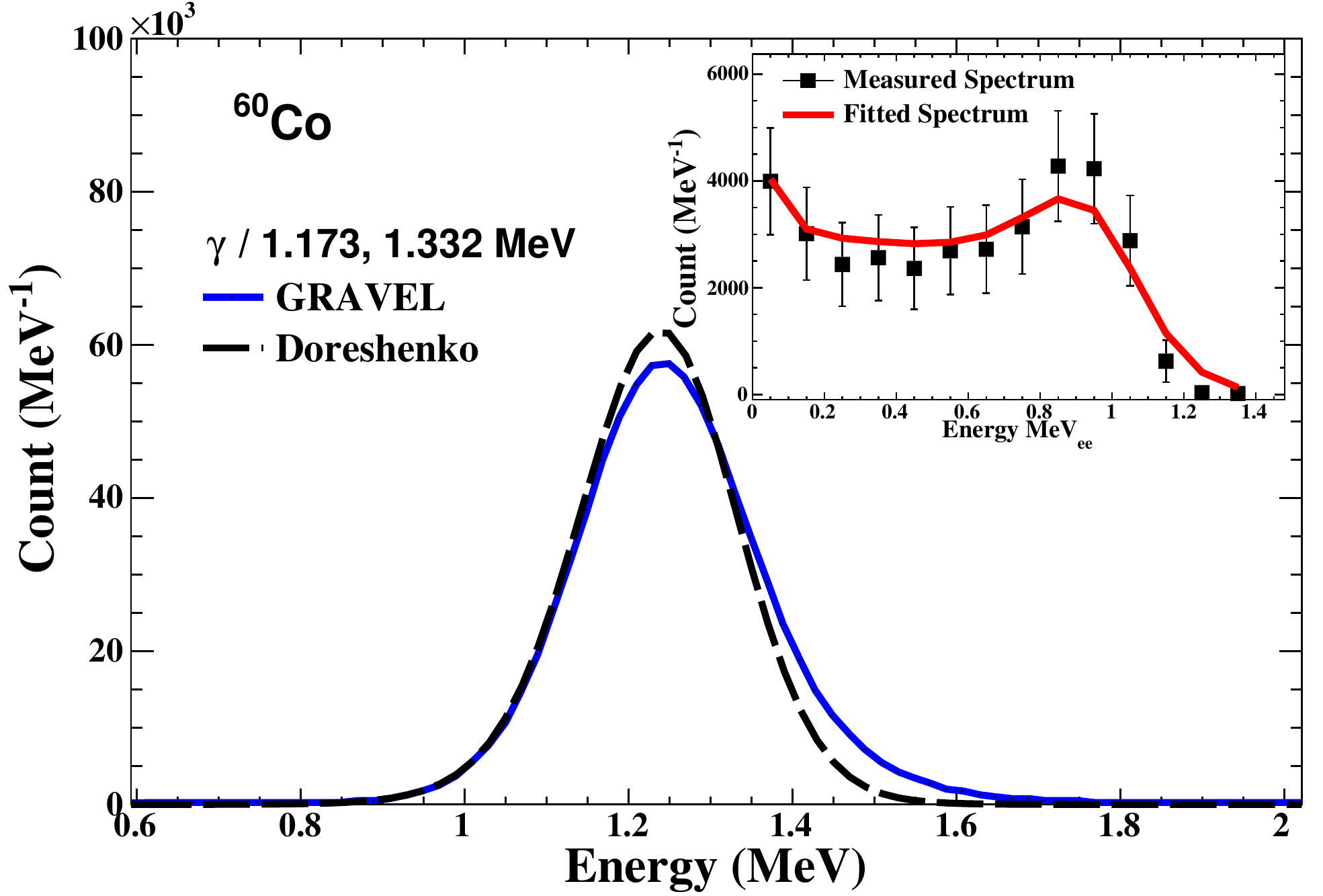}
%\end{minipage}
\caption{(color online) Recoil electron spectra and the unfolded gamma-rays energy spectrum of (a) $^{241}$AmBe, (b) $^{22}$Na, (c) $^{137}$Cs, (d) $^{60}$Co, respectively.}
\label{fig::gamma}
%\end{center}
\end{figure}
This algorithm calculates the $j^{th}$ bin of the neutron flux $\Phi_j(E_j)$. The index $n$ over the $\Phi$ indicates the iteration number. The algorithm works with the inputs $N_i(L_i)$, detector measurement, response function $R_{ij}(L_{i},E_{j})$ and an initial neutron flux $\Phi_j^{0}(E_j)$. This initial flux is just to get the algorithm started, but if there is a strong guess on the expected neutron flux, it may speed up the calculations, or even may result in a more precise result~\cite{vega2002}. In general, a few hundreds of iterations are enough for the calculation to converge, and a good fit can be obtained.

The Gravel unfolding method can be written as,

\begin{equation}
\Phi_{j}^{n+1}=\Phi_{j}^{n}exp\left( \frac{\sum_{i}W_{ij}^{n}ln\left(\frac{N_{i}}{\sum_{k}\Phi_{k}^{n}R_{ik}}\right)}{\sum_{i}W_{ij}^{n}} \right),
\label{eq::gravel}
\end{equation}
where $R_{ij}$ represents the response function obtained from Eq.~\ref{eq::rf}. $N_{i}$ is the measured counts in $i^{th}$ bin light output. $W_{ij}$ is a weight factor defined as,
\begin{equation}
W_{ij}^{n}=\frac{R_{ij}\Phi_{j}^{n}}{\sum_{k}\Phi_{k}^{n}R_{ik}}\frac{N_{i}^{2}}{\sigma^{2}},
\label{eq::wfac}
\end{equation}
where $\sigma$ represents the RMS error of the light output. The convergence of the iterative procedure is controlled by $\chi^{2}/n.o.f$ given in Eq.~\ref{eq::grachi}. When it is close to unity the iteration will be terminated.
\begin{equation}
\chi^{2}/n.o.f = \frac{1}{n.o.f}\sum_{i}\frac{\left(\sum_{j}R_{ij}\Phi_{j}-N_{i}\right)^{2}}{\sigma_{i}^{2}}.
\label{eq::grachi}
\end{equation}

The detector has been exposed to a well known fast neutron source of $^{241}$AmBe($\alpha$,n) to test the unfolding methods. The fast neutron band shown in Figure~\ref{fig::psdDist} which is the recoil spectrum shown in Figure~\ref{fig::ambe}a is taken as an input for both of the iterative methods. The resulting unfolded $^{241}$AmBe($\alpha$,n) fluxes for both methods are shown in Figure~\ref{fig::ambe} along with the measurement of Kluge and Weise in 1982~\cite{kluge82}. As can be seen a perfect match of the peak positions and the shape in general are obtained successfully. Therefore, the simulation package developed in this study is able to reproduce the actual neutron energy spectra from the recoiled spectra. 

In order to illustrate the robustness of our iteration method, the same unfolding procedure as in the reconstruction of actual neutron spectrum can also be applied to obtain the actual gamma spectra of $^{22}Na$, $^{137}Cs$, and $^{60}Co$, which are used for detector energy calibration. The reconstructed actual energy spectra of the gamma sources are demonstrated in Figure~\ref{fig::gamma}. The actual gamma spectra are reconstructed successfully from the measured recoil gamma spectra together with the corresponding gamma response functions given in Figure~\ref{fig::resp_all_gamma}. The gamma peaks are located at the mean of (666.9 $\pm$ 80.6) $keV_{ee}$ for $^{137}Cs$, (1277.8 $\pm$ 93.1) $keV_{ee}$ and (524 $\pm$ 77.8) $keV_{ee}$ for $^{22}Na$ and (1254.7 $\pm$ 105.9) $keV_{ee}$ for $^{60}Co$, respectively.

Our on-going research program is to install this detector at the actual shielding configurations at KSNL \cite{ksnl} to provide ambient neutron background measurements. 

\section{Summary and Conclusion}

Neutrino and Dark Matter experiments with low event rates highly depend on background suppression methods. Neutron component of the ambient background radiation is especially problematic since neutrons are difficult to shield directly. A new research window can be opened for measuring actual neutron flux at the experimental site to measure and therefore to subtract neutron background from the physics signal.

We report in this article our efforts on the optimization and characterization of a HND composed of BC501A liquid and BC702 organic scintillating neutron detector. Monte Carlo Simulation tools are developed with the aim of obtaining actual neutron energy spectra and neutron flux from the measurement of partially scattered neutron spectra. The $^{241}$AmBe($\alpha$,n) neutron energy spectra are reconstructed successfully by means of Doroshenko and Gravel unfolding algorithm. This paves the way for in situ background measurement currently being pursued.

\section{Acknowledgments}

This work is supported by contract 114F374 under TUBITAK, Turkey and 104-2112-M-001-038-MY3 from the Ministry of Science and Technology, Taiwan. M K Singh thanks the University Grants Commission (UGC), Govt. of India, for the funding through UGC D. S. Kothari Postdoctoral Fellowship (DSKPDF) scheme.

\end{document}